\documentclass[aps,pra,10pt,twocolumn,amsmath,citeautoscript,groupedaddress,superscriptaddress]{revtex4-2}

\usepackage[colorlinks,urlcolor=blue,citecolor=blue]{hyperref}
\usepackage[rgb,dvipsnames]{xcolor}
\usepackage[export]{adjustbox}
\usepackage{graphicx}
\usepackage{bbold}
\usepackage{subcaption}
\usepackage{qcircuit}
\usepackage{braket}
\usepackage{amsmath,amssymb}
\usepackage[per-mode=symbol]{siunitx}
\DeclareSIUnit\angstrom{\text{\AA}}

\begin{document}

\title{An assessment of quantum phase estimation protocols for early fault-tolerant quantum computers}

\author{Jacob S. Nelson}
\affiliation{Center for Quantum Information and Control (CQuIC), Department of Physics and Astronomy, University of New Mexico, Albuquerque, NM 87131, USA}
\author{Andrew D. Baczewski}
\affiliation{Center for Quantum Information and Control (CQuIC), Department of Physics and Astronomy, University of New Mexico, Albuquerque, NM 87131, USA}
\affiliation{Quantum Algorithms and Applications Collaboratory (QuAAC), Sandia National Laboratories, Albuquerque, NM 87185, USA}

\date{\today}

\begin{abstract}
We compare several quantum phase estimation (QPE) protocols intended for early fault-tolerant quantum computers (EFTQCs) in the context of models of their implementations on a surface code architecture.
We estimate the logical and physical resources required to use these protocols to calculate the ground state energy of molecular hydrogen in a minimal basis with error below $10^{-3}$ atomic units in the presence of depolarizing logical errors. 
Accounting for the overhead of rotation synthesis and magic state distillation, we find that the total $T$-gate counts do not vary significantly among the EFT QPE protocols at fixed state overlap. 
In addition to reducing the number of ancilla qubits and circuit depth, the noise robustness of the EFT protocols can be leveraged to reduce resource requirements below those of textbook QPE, realizing approximately a 300-fold reduction in computational volume in some cases.
Even so, our estimates are well beyond the scale of existing early fault-tolerance demonstrations.
\end{abstract}

\maketitle

\section{Introduction}
\label{sec:intro}
As quantum computing technologies continue to mature, demonstrations of elementary logical operations have become possible~\cite{ryan2021realization,google2023suppressing,yamamoto2023demonstrating,wang2023faulttolerant,bluvstein2023logical}.
In anticipation of devices with a few logical qubits, a body of work has developed around the design and analysis of algorithms for so-called early fault-tolerant quantum computers (EFTQCs)~\cite{campbell2021early,dong2022ground,ni2023low,li2023low,ding2023even,ding2023simultaneous,ding2024quantum,katabarwa2023early}.
One of the proposed applications of EFTQCs is quantum phase estimation (QPE), but textbook QPE~\cite{Kitaev1995QuantumMA,mikeIke} has very demanding resource requirements and it is relatively sensitive to logical errors.
Thus, much of the work on algorithms for EFTQCs has focused on methods requiring only a single ancilla qubit, reducing circuit depths, and improving robustness to errors~\cite{ni2023low,li2023low,ding2023even,ding2023simultaneous}.
However, relatively little focus has been given to the impact of the overheads associated with implementing quantum error correction (QEC) and fault-tolerant operations~\cite{Litinski2019gameofsurfacecodes}, thus it is presently unclear which approach is actually best suited to EFTQCs.

QEC uses many error-prone physical qubits to encode a smaller number of logical qubits that have a lower effective error rate~\cite{shor1995scheme,calderbank1996good}.
QEC is a component of implementing fault-tolerant quantum computation, in which logical operations are applied to encoded logical qubits in such a way that errors can be corrected at least as quickly as they occur~\cite{shor1996fault}.
However, the resource overheads are significant and ultimately the dominant consideration in determining the suitability of any algorithm for an EFTQC.
As hardware that might be capable of implementing FTQC continues to develop, a careful analysis of the resources required to run specific algorithms is necessary in order to understand not only which applications can be run, but how to make better use of EFTQCs.

Other authors have recently demonstrated the gap between the resource requirements for scalable surface code implementations of QPE and current generations of quantum hardware~\cite{blunt2023compilation}, which are arguably EFT or at least approaching that.
In this manuscript, we analyze QPE protocols designed for EFTQCs in the context of QEC overheads, explicitly accounting for differing degrees of robustness to simple logical errors. 
Our aim is to understand the relative efficiency of these different approaches in the context of the first scalable QPE demonstrations and to highlight the practical costs of fault-tolerant implementation for EFT algorithm developers. 
The earliest EFT QPE demonstrations will be (and are~\cite{yamamoto2023demonstrating}) based on more customized QEC codes with less significant resource requirements and are likely to combine QEC and error mitigation~\cite{suzuki2022quantum}.
However, we will focus entirely on scalable surface code demonstrations that currently seem to be among the best practical candidates for realizing low logical error rates at large distance with a high threshold and straightforward connectivity requirements~\cite{fowler2012surface}.

QPE protocols solve the problem of estimating an eigenphase ($\phi_0$) of a unitary operator ($U$) in which $U\ket{\psi_0}=e^{i\phi_0}\ket{\psi_0}$, given access to a state $\ket{\psi}$ with overlap $\gamma=|\langle \psi_0 | \psi \rangle|^2$~\cite{mikeIke,kitaev2002classical}.
Applications of QPE include ground-state energy estimation, in which $\ket{\psi_0}$ is the ground state of some many-body Hamiltonian ($H$) for a physical system of interest (e.g., interacting spins~\cite{clark2009resource}, molecules~\cite{aspuru2005simulated}, or solids~\cite{pathak2023quantifying}).
If $U$ is the time evolution operator $e^{-iHt}$ and $\ket{\psi_0}$ is the ground state of $H$, then QPE provides an estimate for the ground state energy of $H$.
Other applications of QPE include more general observable estimation and use as a subroutine in more elaborate quantum algorithms~\cite{knill2007optimal,harrow2009quantum,temme2011quantum}, here we focus primarily on energy estimation as an early application~\footnote{Arguably, the use of QPE in metrology~\cite{berry2001optimal,QMetrology,higgins2007entanglement,} and the calibration of gates on physical qubits~\cite{kimmel2015robust,rudinger2017experimental} is earlier, but does not require fault tolerance.}.
While QPE can achieve asymptotically optimal Heisenberg scaling~\cite{berry2001optimal,QMetrology,Zhou_2018Hesienburg,OptimalityHeisenburg}, which is a quadratic improvement over the standard quantum limit (SQL)~\cite{SQL}, the requisite circuits depths are expected to be large enough that textbook QPE protocols will not be feasible even on EFTQCs.
To address this, recent work has been devoted to making QPE protocols better suited for EFTQCs~\cite{dong2022ground,ni2023low,li2023low,ding2023even,ding2023simultaneous,ding2024quantum}.

In this manuscript, we analyze the performance of some of these protocols applied to the problem of estimating the ground-state energy of molecular hydrogen in a minimal basis.
We estimate the resources required to implement this calculation on an FQTC running the surface code, accounting for the impacts of rotation synthesis~\cite{ross2014optimal}, magic state distillation~\cite{kitaevUniversal2005,Litinski2019magicstate}, a nonzero logical error rate, and imperfect state prep in which $\gamma<1$.
We generally find that QPE protocols based on the Hadamard test have comparable overheads, though there is some differentiation among the resource requirements for $\gamma<1$ and low precision.

The contents of this manuscript are as follows.
In Section~\ref{sec:protocols}, we describe the relevant QPE protocols.
In Section~\ref{sec:surface_code_implementations}, we review how these protocols can be implemented in a surface code architecture.
In Section~\ref{sec:trotterized_H2}, we compare the logical resource requirements for the EFT protocols in a specific instance-- QPE on a Trotterized time-evolution unitary describing molecular hydrogen (H$_2$) in a minimal basis.
In Section~\ref{sec:physical_resource_estimates}, we provide physical resource estimates associated with implementing the various QPE protocols for this exemplar.

\section{QPE Protocols}
\label{sec:protocols}

\begin{figure}[ht]
    \centering
    \includegraphics{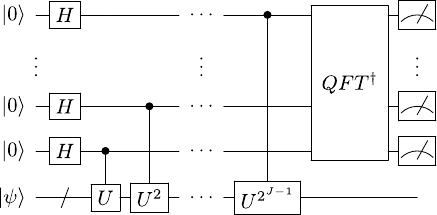}
    \caption{A textbook QPE circuit~\cite{mikeIke}.
    Each ancilla qubit encodes a single bit of an eigenphase of $U$. 
    $H$ indicates a Hadamard gate and $QFT$ indicates the quantum Fourier transform.}
    \label{fig:qpe_circuit}
\end{figure}

A textbook QPE circuit~\cite{mikeIke} is illustrated in Fig.~\ref{fig:qpe_circuit}.

This circuit efficiently computes $\phi_0$ given access to a state $\ket{\psi}$ with overlap $\gamma$. 
However, it is widely regarded as ill-suited for the first FTQCs because it requires many ancilla qubits and deep circuits~\cite{lin2022heisenberg,zhang2022computing}. 
Worse yet, due to the use of the quantum Fourier transform (QFT) textbook QPE is very sensitive to errors and requires a high degree of QEC to ensure sufficiently low logical error rates.
Various alternatives to textbook QPE have been proposed that use one or zero ancilla~\cite{lu2021algorithms,obrien2021error}, reduce the maximum circuit depth~\cite{ding2023even,ding2023simultaneous,li2023low,ni2023low}, or provide robustness to noise or other errors~\cite{kimmel2015robust,russo2021evaluating,ding2023robust}. 
These other protocols do not rely on the QFT and benefit from sampling multiple outcomes rather than estimating the eigenphase in a single shot.
In the remainder of this Section, we describe the four alternatives to textbook QPE that we will consider in this manuscript: iterative phase estimation (IPE) ~\cite{IPE}, robust phase estimation (RPE) ~\cite{ni2023low}, quantum complex exponential least squares (QCELS) ~\cite{ding2023even}, and multi-modal multi-level QCELS (MMQCELS) ~\cite{ding2023simultaneous}.

\subsection{Iterative Phase Estimation}

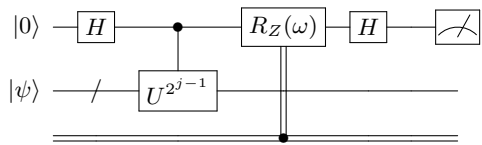
\begin{figure}[ht]
    \centering
    \centerline{
        \Qcircuit @C=1em @R=1em {
        \lstick{\ket{0}} & \gate{H} & \ctrl{1} & \gate{R_Z(\omega)} & \gate{H} & \qw & \meter\\
        \lstick{\ket{\psi}} & {/} \qw & \gate{U^{2^{j-1}}} & \qw \cwx & \qw & \qw & \qw\\
        & \cw & \cw & \control \cw \cwx & \cw & \cw & \cw \\        
        }
    }   
    \caption{An IPE circuit. 
    The inverse $QFT$ is performed semiclassically as a sequence of classically controlled $Z$ rotations, $R_Z(\omega)$. 
    This only requires a single ancilla qubit.}
    \label{fig:ipe_circuit}
\end{figure}

The inverse QFT in Fig.~\ref{fig:qpe_circuit} can be effected semiclassically ~\cite{PhysRevLett.76.3228} using a single ancilla and classically controlled rotations. 
This is leveraged in IPE to reduce the number of ancilla qubits to 1. In IPE the circuit shown in Fig.~\ref{fig:ipe_circuit} is run for $J$ iterations to extract $J-1$ bits of the phase.

The classically controlled rotation is conditioned on the measurement outcomes of the previous iterates. 
While the ancilla is measured and reset after each iterate, the system register can remain coherent across iterates. 
If $\gamma=1$ then in the absence of errors the probability of successfully extracting $J-1$ bits using $J$ iterates is always greater than $8/\pi^2$, independent of $J$~\cite{IPE}. 
In practice, errors and state preparation for which $\gamma<1$ are likely to be a concern, especially in the EFT regime. 
As such, we consider the effects of imperfect state preparation and depolarizing errors. 
While the errors in actual quantum processors are likely to be more complicated~\cite{sarovar2020detecting}, a depolarizing error model provides some insight into the effects of errors while remaining analytically tractable. 

The probability of successfully measuring the correct value of $J-1$ bits is bounded from below as
\begin{equation}
    \label{eq:ipe_success_ps}
    p_s(\gamma) \geq \frac{\gamma}{2} \prod \limits_{j=1}^{J-1}\left(1+e^{-a_{j}}\cos\left(\frac{\pi}{2^{j+1}}\right)\right),
\end{equation}
where $a_j$ quantifies the aggregate effect of depolarization in the $j$th iterate (See Appendix~\ref{app:error_model}). 
To overcome the effects of depolarization and achieve $p_s \approx 1$ the logical qubits in our computation will have to be encoded at a sufficient distance $d$ to realize a low target logical error rate, the details of which will be discussed in Section~\ref{sec:physical_resource_estimates}. 

\subsection{Robust Phase Estimation}

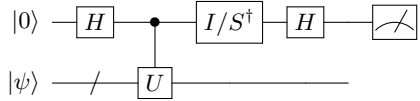
\begin{figure}[ht]
    \centering
    \centerline{
        \Qcircuit @C=1em @R=1em {
        \lstick{\ket{0}} & \gate{H} & \ctrl{1} & \gate{I/S^{\dagger}} & \gate{H} & \qw & \meter\\
        \lstick{\ket{\psi}} & {/} \qw & \gate{U} & \qw & \qw & \qw \\
        }
    }
    \caption{The Hadamard test circuit.}
    \label{fig:Had_test}
\end{figure}

Perhaps the simplest algorithm for phase estimation is the Hadamard test shown in Fig.~\ref{fig:Had_test}.
Contrasting this with the semiclassical QFT in IPE, we only have to perform a single controlled unitary.
The $I/S^{\dag}$ gate represents separate instances of the circuit where either $I$ or $S^{\dag}$ is applied. 
Applying the $I$ ($S^{\dag}$) gate yields a sample of $x$ ($y$). 
When $\ket{\psi}$ is an eigenstate of $U$ with eigenvalue $e^{-i\phi_0}$ the expected values $\langle x \rangle=\cos{\phi_0}$ and $\langle y \rangle = \sin{\phi_0}$, such that atan2$(\Bar{x},\Bar{y})$ is used to estimate the phase, where $\Bar{x}$ ($\Bar{y}$) indicates the sample mean.

While straightforward, the Hadamard test has the drawbacks of requiring $\ket{\psi}$ to be an exact eigenstate and having SQL scaling $N_s = \mathcal{O}(\epsilon^{-2})$~\footnote{In contrast to the optimal Heisenberg-limited scaling $N_s = \mathcal{O}(\epsilon^{-1})$.}. 
RPE allows for $\gamma<1$ and achieves Heisenberg scaling by adding an additional parameter, the number of iterates $J$. 
This requires modifying the Hadamard test circuit as follows
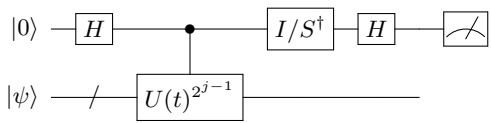
\begin{figure}[ht]
    \centering
    \centerline{
        \Qcircuit @C=1em @R=1em {
        \lstick{\ket{0}} & \gate{H} & \ctrl{1} & \gate{I/S^{\dagger}} & \gate{H} & \qw & \meter\\
        \lstick{\ket{\psi}} & {/} \qw & \gate{U(t)^{2^{j-1}}} & \qw & \qw & \qw 
        }
    }
    \caption{Hadamard test circuit for $U(t)^{2^{j-1}}$.}
    \label{fig:H_test}
\end{figure}

The circuit is iterated over integer values of $j$ from $1$ to $J$.
Applying the time-evolution operator $2^{J-1}$ times leverages the Heisenberg scaling with $t$ such that the error scales like $2^{-(J-1)}$.
For each $j$, atan2 is used to estimate $\theta_j$.
The search interval for the next estimate is then halved such that $\theta_{j+1} \in [\theta_j-\frac{\pi}{2^{j}}, \theta_j+\frac{\pi}{2^{j}}] \mod{2\pi}$. 
Thus, when compared to the Hadamard test, the error on the $J$th estimate is reduced by $\frac{\pi}{2^{J-1}}$, provided the error on each $\theta_j$ does not exceed $\frac{\pi}{2^{j}}$. 
Using Hoeffding's inequality we can ensure this condition is met with high probability as long as the $\gamma \geq 0.536$~\cite{ni2023low}.

We note that the estimator for RPE becomes biased  for $\gamma < 1$, however, depolarization does not contribute to the bias. 
Considering depolarizing noise and imperfect state preparation the error on the final estimate is
\begin{equation}
    \epsilon = \frac{1}{t2^{J-1}}\sqrt{\frac{e^{2a_J}}{N_s}+\eta^2(\gamma)},
\end{equation}
where $\eta(\gamma)$ is the bias, which is upper bounded in Appendix~\ref{app:error_model}.
We further note that because each circuit only implements a single controlled $U^{2^{j-1}}$, there is a constant factor reduction in depth when compared to textbook QPE.

\subsection{Quantum Complex Exponential Least Squares}

QCELS utilizes circuits of the same form as Fig.~\ref{fig:H_test}, i.e., the Hadamard test. 
It relies on a complex random variable $z(t) = \Bar{x}(t) + i\Bar{y}(t)$. When $\ket{\psi}$ is an exact eigenstate the expected value of $z(t)$ is $e^{-i\lambda t}$, where $\lambda$ is related to the energy eigenvalue. 
At fixed $j$, the circuit is run for $N$ values of $t$. 
The resulting data will take the form of a complex exponential with frequency $\lambda$. 
A least-squares fit is then applied assuming this model and the best-fit parameter $\hat{\lambda}_j$ is output as the estimate for  $\lambda$. 
This routine is then repeated for all values for $j$, with the previous estimate $\hat{\lambda}_{j}$ used to seed the least-squares fit for $(j+1)$th round.

When $\gamma<1$, the resulting data will necessarily contain frequency contributions from other eigenenergies. 
In this case the least-squares fit will attempt to extract the dominant frequency component. 
Provided that the $\gamma\geq 0.71$ the protocol will succeed with high probability. 
We note that the data point at $t = 0$ can be inferred as $e^{0}=1$. 
In order to prevent the query complexity from exceeding that of RPE only 1 additional data point should be taken (i.e., $N=2$).
Implemented in this manner, QCELS and RPE differ only in post-processing.
Thus QCELS achieves the same error as RPE at equivalent circuit depths, though post-processing details can be nontrivial~\cite{russo2021consistency}. 
Numerically, we found that the bias due to imperfect state prep is the same for QCELS as it is for RPE. 

A few variations on QCELS have also recently been published. 
The first of which introduces a multi-modal version of the algorithm, dubbed multi-modal QCELS (MMQCELS). 
In MMQCELS the Hamiltonian simulation times are randomly drawn from a normal distribution and the fitting subroutine can output estimates for multiple eigenvalues at once. 
When applied to single eigenvalue estimation, MMQCELS retains the same error as QCELS, while permitting $\gamma \geq 0.5$ ~\cite{ding2023simultaneous}. 
We include $T$-gate counts for both QCELS and MMQCELS in Sec.~\ref{sec:trotterized_H2}. 
Further work proposed including knowledge of the errors/noise into the fitting subroutine to increase robustness~\cite{ding2023robust}.
While we do not investigate this proposal here, we note that it would be interesting to study in the same context as this work.

\section{Surface Code Implementations}
\label{sec:surface_code_implementations}

In this Section we review the model according to which we quantify the space and time overheads associated with implementing these protocols in a surface code architecture.
We rely entirely on the lattice surgery implementation described by Litinski to model these costs~\cite{Litinski2019gameofsurfacecodes,Litinski2019magicstate}.

The space overhead associated with a surface code implementation of any quantum algorithm is driven by the number of physical qubits required to encode the requisite number of logical qubits at a sufficiently low logical error rate.
While it is a potentially severe approximation, we ascribe a single monolithic error rate of $p$ to all physical qubits.
Logical qubits are comprised of many physical qubits encoded in a QEC code and are ideally less prone to errors. 
Here we will focus on the surface code, for which $2d^2$ physical qubits are required to encode a distance-$d$ logical qubit, where $d$ is twice the number of physical errors that can be corrected per logical qubit~\cite{dennis2002topological}.
We also assign a monolithic error rate to each logical qubit of $p_L$, which is a function of $p$ and $d$ (\emph{vide infra} Eq.~\ref{eq:logical_error_model}).

The temporal overhead is quantified by the number of QEC code cycles required to implement a particular quantum computation. 
Simply put, a code cycle is a single discrete step in a computation on an FTQC.
The actual physical time associated with implementing a code cycle will depend on the specific hardware platform and architecture.
We will primarily focus on quantifying the resource requirements in terms of the number of ``hard'' non-Clifford operations required to implement QPE.

Operations on logical qubits can be categorized as ``easy'' or ``hard'' depending on whether they're transversal (easy) or not (hard). The Eastin-Knill theorem states that transversal gate sets cannot be universal~\cite{eastin2009restrictions}.
Thus any universal gate set is necessarily partitioned into easy and hard operations.
For the surface code, easy operations include gates within the Clifford group, whereas non-Clifford gates are hard to implement. 
Any unitary can be approximated to arbitrary accuracy as sequences of Cliffords and non-Clifford $T$ gates~\cite{Dawson2005TheSA} and thus we analyze the resource requirements in terms of the number of $T$ gates. 
In fact, the Clifford gates don't need to be directly implemented at all. 
Using commutation relations, all Clifford gates can be propagated through to the end of the circuit, where their action is to change the basis in which measurements are performed~\cite{Litinski2019gameofsurfacecodes}. 
By propagating these operations through the circuit, the final measurements can be performed in the transformed bases instead of directly implementing the Clifford gates.
The non-Clifford gates are transformed by this process and are now described as Pauli product rotations by angle $\pi/8$.
The number of non-Clifford operations is conserved in this process and requires the same physical resources to implement.

\begin{figure}[ht]
    \centerline{
        \Qcircuit @C=0.6em @R=0.75em {
        \lstick{\ket{\psi}} &{/} \qw &\multimeasure{2}{M_{PZ}}& \gate{R_P(\pi/2)} & \gate{R_P(\pi)} & \qw & \rstick{R_P(\pi/4)\ket{\psi}}\\
         & &\push{\rule{3.3em}{0em}}& \control \cw  \cwx \\
        \lstick{\ket{T}} & \qw& \ghost{M_{PZ}} & \qw & \measure{M_X} \cwx[-2] \\
        }
    }
    \caption{A circuit in which a $T$-like operation is applied via gate teleportation.
    Note that $M_X$ refers to an $X$ measurement and $M_{PZ}$ refers to a generalized Pauli $P \otimes Z$ measurement, where $P$ is the Pauli string leftover after commuting away Clifford operations as described in the text and Ref.~\cite{Litinski2019gameofsurfacecodes}.}
    \label{fig:T-Op}
\end{figure}
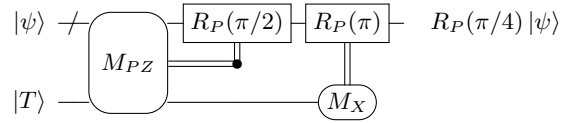

The non-Clifford ``$T$-like'' gates can not be implemented transversally within the surface code.
Instead these operations are implemented by performing a joint measurement with a $T$ state $\ket{T} = 1/\sqrt{2}\left(\ket{0}+e^{-i\pi/4}\ket{1}\right)$ and applying a correction based on the measurement outcome~\cite{Litinski2019magicstate}.
Explicit construction of this operation is given in Fig.~\ref{fig:T-Op}.
These $T$ states cannot be be prepared fault-tolerantly and must instead be prepared via $T$ state distillation.
The quality and rate of $T$ state production depends on the protocol chosen for state distillation.
For example, the 15:1 distillation protocol uses 11 logical qubits and distills $T$ states with error rate $35p^3$ once every $11d$ code cycles.

In addition to the logical qubits that are used for $T$-state distillation, we require logical qubits that are used for routing as well as logical qubits that represent the computational degrees of freedom in any given algorithm (e.g., the ``computational'' qubits indicated in the circuit diagrams in Sec.~\ref{sec:protocols}).
We focus on Litinski's compact set up, which uses the minimum number of logical qubits~\cite{Litinski2019gameofsurfacecodes}.
The compact block requires $\lceil1.5n+3\rceil$ logical qubits to encode $n$ computational qubits and can implement one $T$ gate every $9d$ code cycles.
More qubits can be used to perform the $T$ gates faster.

The total number of logical qubits is then determined by the number needed to encode the computational qubits and the number needed to distill $T$ states.
As indicated previously, the number of physical qubits per logical qubit for a surface code of distance $d$ is $2d^2$.
The number of code cycles to perform a single $T$ gate is then the maximum of the number of code cycles required for the distillation protocol producing $T$ states and the number of code required for the computational qubits to consume them, which we indicate as $T_{rate}$.

We now define two key metrics when considering logical circuits to be run on a surface code.
The maximum number of $T$ gates in any one circuit $T_{max}$ dictates the maximum coherence time needed and informs the choice of $d$.
An estimate of the total run-time is obtained from the total number of $T$ gates $T_{tot}$.
In total, the number of code cycles to run the full computation will be $T_{rate}T_{tot}d$.

\section{Trotterized H$_2$ Simulations}
\label{sec:trotterized_H2}

We apply each protocol to the estimation of the ground state energy of molecular hydrogen (H$_2$) in a minimal basis set with a Bravyi-Kitaev encoding ~\cite{O_Malley_2016molecules}. 
The corresponding Hamiltonian is then
\begin{equation}
    H = g_1Z_1+g_2Z_2+g_3X_1X_2+g_4Y_1Y_2.
\end{equation}
The time evolution operator $U(t)=e^{-iHt}$ is approximated using a first-order Trotter product formula
\begin{equation}
    U(t) = e^{-ig_1Z_1t}e^{-ig_2Z_2t}e^{-ig_3X_1X_2t}e^{-ig_4Y_1Y_2t}.
\end{equation}
Each term in the unitary is thus a generalized Pauli rotation that can be implemented with Clifford gates and a single $Z$ rotation. 
In order to make the unitary controlled, we need only add controls to the $Z$ rotations. 
We perform the slight optimization of utilizing directional control ~\cite{WeckerDirectControl} to implement the controlled rotations as an uncontrolled rotation and two CNOTS. 
Our controlled unitary thus costs 4 $Z$ rotations per query. 
We show explicit circuits in Appendix~\ref{app:circuit_constructions}

At the beginning of every QPE circuit we must first prepare the system register in the ground state of $H$. 
While there exist many algorithms for ground state preparation~\cite{farhi2000quantum,poulin2009preparing}, investigation of these protocols falls outside the scope of this work. 
Instead, as the system only requires 2 qubits, we find the eigenstates exactly and construct a projector from the state $\ket{00}$ to the exact ground state. 
To study the dependence of the QPE protocols on $\gamma$ we simply apply a rotation to one of the system qubits before applying the projector, allowing us to assign a particular value to $\gamma$.

We test each protocol at a target error $\epsilon$ spanning 3 orders of magnitude. 
For each $\epsilon$, we construct all of the circuits required by each protocol to achieve error $\epsilon$ with success probability $P_s > 0.99$ using only Clifford gates and $Z$ rotations. 
We then use the \textsc{gridsynth} software package to decompose the $Z$ rotations into sequences of 1-qubit Clifford and $T$ gates ~\cite{ross2014optimal}. 
Since we expect rotation synthesis error to grow linearly (at worst) with the number of $Z$-rotations, we synthesize each rotation to an accuracy below the target error divided by the total number of $Z$ rotations.
Other sources of error, e.g., due to Trotterization, will add in quadrature and thus we set an overall error budget that keeps the aggregate error below some target.
For all of our tests we precompute the Trotter error to verify that it is within our overall error budget (see Appendix~\ref{app:circuit_constructions} for more details).
For all protocols we consider $\gamma=1$ and $\gamma=0.75$, which are both above the threshold overlaps for RPE and QCELS~\cite{ni2023low,ding2023even}.

We count the total number of $T$ gates $T_{tot}$ required for each protocol, including all circuits and shots, to achieve the target error. 
We also report the maximum number of $T$ gates $T_{max}$ in any one circuit of each protocol. 
Additionally, we use qiskit's Qasm simulator~\cite{Qiskit} to simulate all circuits to verify the estimates from each protocol do not exceed the target error. 
Results from the simulated protocols are given in Appendix~\ref{app:numerical_tests}.

\begin{figure}[ht]
\centerline{
\includegraphics[width=0.9\columnwidth]{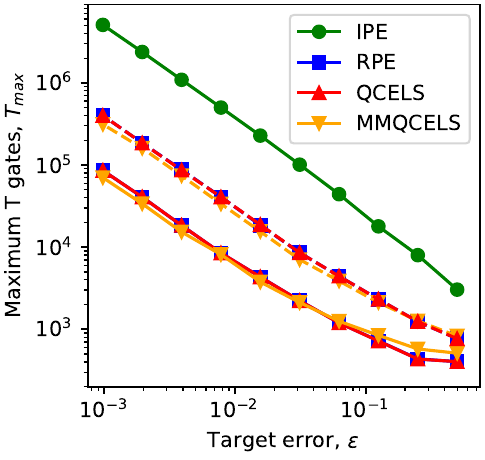}
}
\caption{The maximum number of $T$ gates required in the deepest iteration for each protocol ($T_{max}$) to reach target error $\epsilon$. 
Solid (dashed) lines are for overlap $\gamma=1$ ($\gamma=0.75)$.
Whether $\epsilon$ is realized is verified through numerical simulations of the circuit, which might outperform the indicated value of $\epsilon$.
For the Hadamard test based protocols, reductions in $T_{max}$ are possible at $\gamma=1$ (relative to $\gamma=0.75$) thanks to the ability to reduce the mean-squared error by choosing $N_s$ according to Hoeffding's inequality.
\label{fig:tmax}
}
\end{figure}

\begin{figure}
    \centering
    \includegraphics[width=0.9\columnwidth]{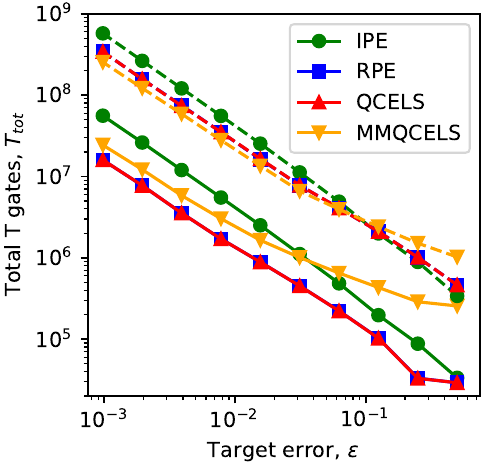}
    \caption{The total number of $T$ gates required across all iterations for each protocol ($T_{tot}$) to reach target error $\epsilon$.
    Solid (dashed) lines are for overlap $\gamma=1$ ($\gamma=0.75)$.
    That we are achieving an error of at most $\epsilon$ is verified through numerical simulations of the circuit (see Appendix~\ref{app:numerical_tests}). \label{fig:ttot}}    
\end{figure}

We note that for small enough $\epsilon$, $T_{tot}$ does not vary significantly between protocols. 
As RPE, QCELS, and MMQCELS each decompose the controlled unitaries into many Hadamard test circuits they provide a constant factor reduction in $T_{max}$ over IPE.
We note that the $T$ counts for IPE are based off of a worst-case analysis.
The numerics in Appendix~\ref{app:numerical_tests} indicate that IPE tends to outperform this worst case bound and we generally expect that it can be made to be more competitive with the Hadamard test protocols. 

\section{Physical Resource Estimates}
\label{sec:physical_resource_estimates}

We now consider the overheads required to achieve $\epsilon < 10^{-3}$ with $P_s > 0.99$ using textbook QPE, IPE and RPE on a fault-tolerant quantum computer in the presence of depolarizing errors using lattice surgery operations on the rotated surface code. 
We expect the resource requirements for all protocols based on the Hadamard test (RPE, QCELS, and MMQCELS) to be similar, and thus we only compare them for IPE and RPE.

Given a physical error rate per gate of $p$, the logical error rate per code cycle of a surface code with distance $d$ can be approximated as~\cite{fowler2019low}
\begin{equation}
    p_L(p,d) \approx 0.1n_L(100p)^{\frac{d+1}{2}},\label{eq:logical_error_model}
\end{equation}
where $n_L$ is the number of logical qubits. 

IPE and RPE each use $n = 3$ computational qubits. 
The compact setup requires $\lceil 1.5n+3 \rceil$ logical qubits and can consume 1 $T$ state every $9d$ code cycles. 
Using the compact setup thus requires 8 logical qubits. 
We can use the 15:1 distillation protocol protocol to distill a $T$ state with infidelity $b=35p^3$ every $11d$ code cycles using 11 logical qubits. 
Thus using the compact setup along with one 15:1 $T$ state factory would take 19 logical qubits in total and $11dT_{tot}$ code cycles.

We assume a physical depolarization rate of $p = 10^{-3}$, this implies that the $T$ states distilled from the 15:1 protocol will have infidelity $3.5 \times 10^{-8}$.
Consuming a depolarized $T$ state results in noise in the system that is more structured than depolarization. 
For example applying a $T$ gate to a single qubit by consuming a slightly depolarized $T$ state introduces dephasing~\cite{fazio2024logical}, decaying the coherences of the qubit while not affecting the populations. 
With more general multi-qubit non-Clifford operations, the complexity of the error model will grow.
We argue that for single-ancilla protocols this noise can be approximated as depolarization. 
Even the shallowest of circuits considered contain $>100$ $T$ operations with the deepest circuits requiring upwards of $10^6$. 
We expect that in the multitude of these operations, the structure of these errors will be partially washed out and appear random. 
This is compounded by single-ancilla protocols only requiring measurement of a single qubit per circuit.
Under this assumption, consuming $T$ states that have depolarized by amount $b$ will cause the same amount of depolarization on the qubits to which the gate is applied. 

The aggregate error rates $a_j$ (see Section~\ref{sec:protocols}) express the amount of depolarization at the algorithmic level in terms of the physical depolarization rate, QEC code distance, and $T$ state infidelity as
\begin{equation}
    a_j \approx T_{j}(b+T_{Rate}n_Ld0.1(100p)^{\frac{d+1}{2}}),
    \label{eq:aj}
\end{equation}
where $T_j$ is the number of $T$ gates in the $j$th iterate of the protocol. 
We note that for RPE $T_J = T_{max}$. 
As shown in Eq.~\ref{eq:aj}, increasing the code distance decreases the algorithmic noise rate assuming the physical noise is below threshold, however as the $T$ gates are not implemented fault tolerantly, increasing the code distance does not suppress noise originating from the $T$ states. 
The only way to reduce that noise is to use more expensive distillation protocols. 
We note that there is a trade off between increasing code distance and using more expensive $T$ state distillation.
For the problem of H$_2$ considered here, the distillation protocol contributes significantly to the space requirements of the quantum computer. 
In this case, the savings of using less expensive distillation protocols outweighs the costs of slightly increasing the code distance. 

We can now find the minimum code distance required to achieve $\epsilon < 10^{-3}$ using Eq.~\ref{eq:ipe_success_ps}. 
For RPE, we use Hoeffding's inequality to select $N_s$ such that we have a success probability greater than 0.99. 
Using the 15:1 distillation protocol, these conditions can be satisfied for RPE at both of the values of $\gamma$ that we consider. 
For $\gamma=1$, the SQL scaling with $N_s$ can be leveraged to reduce $J$ all the way down to 1 (in which case the protocol is equivalent to the Hadamard test). 
The minimum code distance for $\gamma = 1$ is found to be $d=17$ without reducing $J$ and can be as low as $d=11$ by reducing $J$. 
At $\gamma = 0.75$ the bias is so large that no such reduction in $J$ is possible. 

\begin{figure}[!ht]
    \centering
    \includegraphics[width=0.9\textwidth]{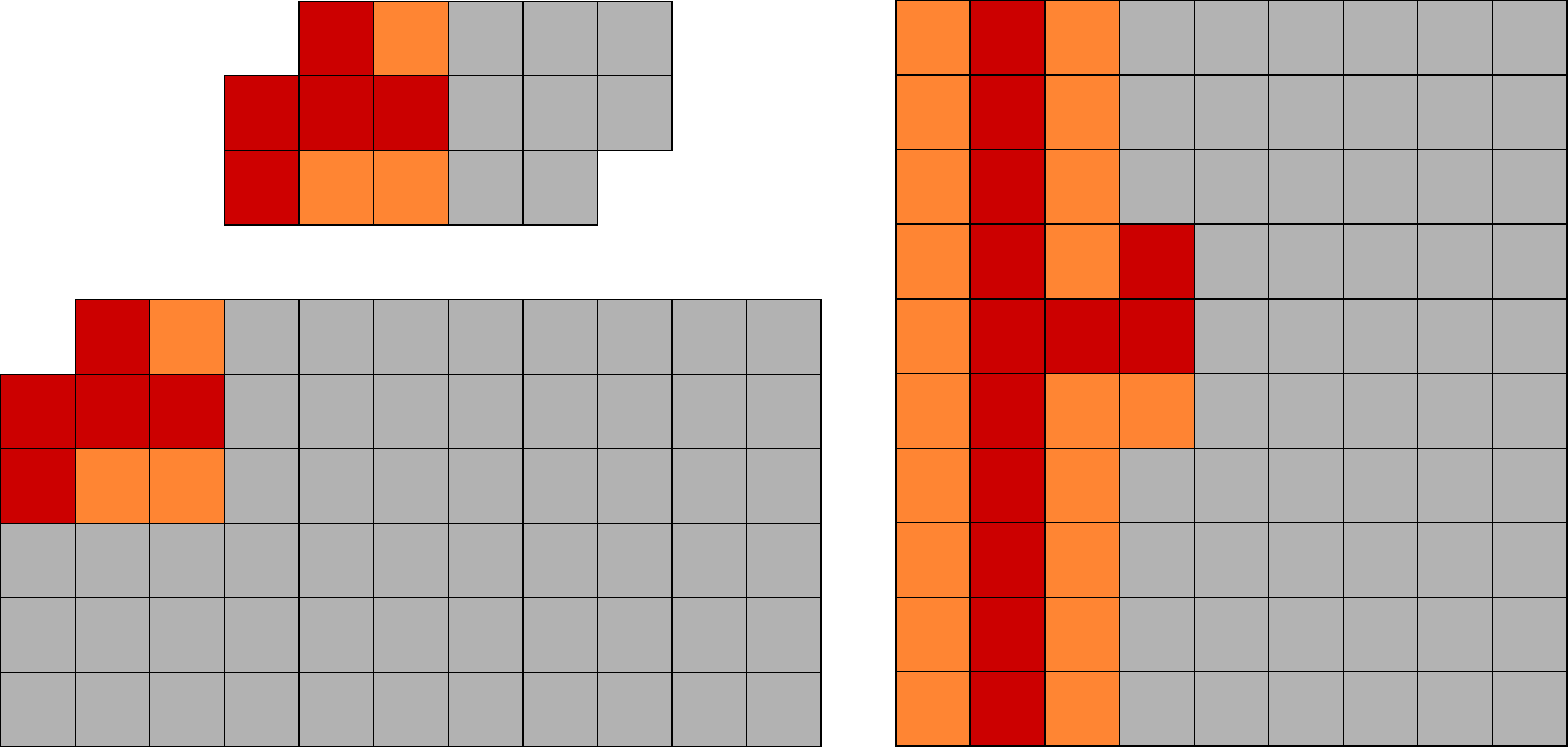}
    \caption{Logical layout for RPE (upper left), IPE with $\gamma=0.75$ (lower left) and textbook QPE (right). At $\gamma=1$, IPE uses the same logical setup as RPE. Orange squares indicate computational qubits, red squares represent routing qubits ensuring proper space for all operations to be performed on the computational qubits, gray squares indicate the space required for T-state factories.}
    \label{fig:qpe_layouts}
\end{figure}

Due to the equivalence in T-counts, we expect the logical layout for RPE in Fig.~\ref{fig:qpe_layouts} to work for QCELS and MMQCELS. 
Additionally, numerical simulations QCELS and MMQCELS to perform similarly to RPE at the same level of depolarizing noise~~\footnote{Source code for our simulations is available at \href{https://github.com/adbacze/qpe4earlyftqc}{https://github.com/adbacze/qpe4earlyftqc}.}.
This is unsurprising as the difference between these protocols lies in the classical post-processing not in the quantum circuits implemented.

For IPE we seek a code distance for which the single-shot probability of success is greater than 0.5 such that the success probability can amplified arbitrarily by increasing the number of shots and taking a majority vote among outcomes. 
At $\gamma = 1$ and $p=10^{-3}$ this can be satisfied using the 15:1 distillation protocol using the logical layout shown in~\ref{fig:rpe_layout} with $d = 21$.
However, for $\gamma = 0.75$ and $p=10^{-3}$ the infidelity of the distilled $T$ states will be too large, requiring the more expensive 116:12 protocol. 
This protocol uses 57 logical qubits, increasing the number of logical qubits up to 65. 
The minimum code distances for this setup was found to be to be $d=21$.

Textbook QPE is extremely sensitive to noise, with even a single error able to cause the protocol to fail. 
As such the distillation protocol and code distance must be set such that with 99\% probability no error occurs.
Additionally, to achieve $\epsilon < 10^{-3}$, QPE requires extra computational qubits. 
This translates to needing 90 logical qubits and a code distance of 27.
Logical qubit layouts are shown for each protocol in Fig.~\ref{fig:qpe_layouts}.

\onecolumngrid

\twocolumngrid

We can now estimate the spacetime overheads required for running each protocol. 
Resource estimates are shown in Table~\ref{tab:resources}.

\begin{table}[ht]
\begin{center}
\begin{tabular}{ |c|c|c|c| } 
 \hline
       Physical Qubits & Protocol & State Overlap ($\gamma$) & Code Cycles \\ 
 \hline \hline
 131,220 & QPE & 0.75 & 1.4$\times 10^{12}$  \\ 
 \hline
 57,330 & IPE & 0.75 & 6.6$\times 10^{11}$   \\ 
 \hline
 10,982 & RPE & 0.75 & 5.0$\times 10^{10}$   \\ 
 \hline
 131,220 & QPE & 1 & 8.4$\times 10^{10}$   \\ 
 \hline
 16,758 & IPE & 1 & 1.4$\times 10^{11}$   \\
 \hline
 10,982 & RPE & 1 & 2.8$\times 10^{9}$   \\ 
 \hline
 4,598 & RPE$^\dagger$ & 1 & 2.1$\times 10^{11}$  \\ 
 \hline
\end{tabular}
\caption{Physical resource estimates for computing the ground state energy of $H_2$ to precision $10^{-3}$. All entries represent a setup using minimal physical qubits. All computations can be sped up by using more qubits to reduce the number code cycles. $^\dagger$There are two entries for RPE at $\gamma=1$ highlighting the trade off between circuit depth and sampling complexity at high overlap. \label{tab:resources}}
\end{center}
\end{table}

Our findings indicate that there exist savings in both physical qubits and code cycles when using IPE or RPE over QPE. 
Beyond requiring fewer ancilla qubits, we were able to leverage robustness to noise to lower the error correction requirements, further reducing the physical overheads. 
Both IPE and RPE possess some tolerance to depolarizing errors.
However, fewer errors build up in RPE due to the constant-factor reduction in circuit depth from splitting up the controlled unitaries across many circuits. 
We note that while we do not provide separate estimates for QCELS and MMQCELS, we expect them to be similar to RPE due to the similarity in circuits implemented.
As previously noted in Section~\ref{sec:trotterized_H2}, the IPE numbers are based upon a worst-case analysis.
The estimates displayed in Table~\ref{tab:resources} could likely be improved in practice.

In order to exploit noise robustness, an assumption of the noise model (in this case pure depolarization) was necessary. 
Realistic errors are likely to be more complicated than depolarization~\cite{rudinger2019probing,proctor2020detecting,sarovar2020detecting}.
A topic of future analysis might be an assessment of the impact of realistic logical errors on algorithm performance.

\section{Conclusions}
\label{sec:conclusions}

We have analyzed the costs of surface code implementations of several alternatives to textbook QPE that have been suggested to be better suited to EFTQCs.
When compiled into logical circuits we find that the total $T$-gate count does not vary significantly between protocols at constant state overlap. 
RPE, QCELS, and MMQCELS provide similar constant-factor reductions to the maximum number of $T$ gates in any given iteration ($T_{max}$).
For $\gamma \approx 1$, $T_{max}$ can be reduced by running fewer distinct iterations at the cost of increasing $T_{tot}$ by running more shots per iteration. 
These trade-offs vanish rapidly as $\gamma$ decreases away from $1$ due to the effects of bias.
The cost of IPE is evidently between the (low) costs of the approaches based on the Hadamard test and (high) cost of textbook QPE.
However, a direct comparison of $T_{tot}$ can be a bit misleading, thanks to the different criteria by which the probability of success is bounded in IPE and Hadamard-based QPE protocols.
IPE tends to overperform in accuracy relative to conservative bounds and it is likely that physical resource requirements could be reduced to be superior to Hadamard-based methods, particularly with the use of Bayesian inference~\cite{wiebe2016efficient}.

Ongoing work is aimed at investigations of other Hamiltonian simulation methods (e.g., QPE on qubitized walks~\cite{Low2019Qubitization}) and implementations on different error correcting code (e.g., 3D color codes~\cite{landahl2014quantum,wang2023faulttolerant}).

\begin{acknowledgements}
We thank 
Zhiyan Ding,
Katerina Gratsea,
Andrew Landahl,
Lin Lin,
Ben Morrison,
Shivesh Pathak, 
and Nathan Wiebe for helpful technical discussions.
We would like to thank the UNM Center for Advanced Research Computing, supported in part by the National Science Foundation, for providing the research computing resources used in this work.
We are grateful to Setso Metodi for pre-publication review and broader administrative support.
All authors were partially supported by Sandia National Laboratories' Laboratory Directed Research and Development (LDRD) Project No.\ 222396 and the DOE Office of Fusion Energy Sciences ``Foundations for quantum simulation of warm dense matter'' project.
Part of this research was performed while the authors were visiting the Institute for Pure and Applied Mathematics (IPAM), which is supported by the National Science Foundation (Grant No. DMS-1925919).

This article has been co-authored by employees of National Technology \& Engineering Solutions of Sandia, LLC under Contract No. DE-NA0003525 with the U.S. Department of Energy (DOE). The authors own all right, title and interest in and to the article and are solely responsible for its contents. The United States Government retains and the publisher, by accepting the article for publication, acknowledges that the United States Government retains a non-exclusive, paid-up, irrevocable, world-wide license to publish or reproduce the published form of this article or allow others to do so, for United States Government purposes. The DOE will provide public access to these results of federally sponsored research in accordance with the DOE Public Access Plan \url{https://www.energy.gov/downloads/doe-public-access-plan}.
\end{acknowledgements}

\appendix 

\section{Error Model}
\label{app:error_model}
We use an error model from Ref.~\cite{liang2023modeling} to analyze the performance of each protocol in the presence of simple depolarizing logical errors. 
For the EFT protocols considered in this work, we are only concerned with the aggregate impact of logical errors on the measurement probabilities of the single ancilla qubit. 
In the error-free case, the probability of measuring $z \in \{0,1\}$ on that qubit is
\begin{equation}
    P(z) = \frac{1}{2}\left( 1+z\cos{\theta} \right).
\end{equation}
For the $j$th iterate of the protocols under consideration, a depolarizing channel with error rate $a_j$ replaces the input state with the maximally mixed state with probability $a_j$,
\begin{equation}
    \mathcal{D}(\rho) = (1-a_j)\rho + a_j\mathbb{1}.
\end{equation}
The probability of remaining error free is then $1-a_j$ resulting in an equivalent decay in the signal.
For small $a_j$, $1-a_j \approx e^{-a_j}$ and the measurement probabilities become
\begin{equation}
    P(z) = \frac{1}{2}\left( 1+z e^{-a_j}\cos{\theta} \right),
\end{equation}

\subsection{IPE Success Probability}
IPE extracts 1 bit of the phase per iterate starting with the $J$th bit. 
For $\gamma=1$, if the phase consists of exactly $J$ bits then the $J$th bit is extracted with success probability $p_J = 1$ in the absence of noise. 
Otherwise $p_J$ will depend on the remainder $\delta$ after $J$ bits of phase. 
In the presence of depolarizing noise, the probability of successfully extracting the $J$th bit is then
\begin{equation}
    p_J(\delta) = \frac{1}{2}\left( 1+e^{-a_J}\cos{2\pi\delta}\right).
\end{equation}

The next iterate attempts to measure the $(J-1)$th bit. 
If all previous bits were correct, then the classically controlled rotations remove all measured bits from the remainder of the current iterate.
The total probability of success for extracting all $J$ bits correctly is
\begin{equation}
    p(\delta) = \prod_{j=1}^J\frac{1}{2}\left(1+e^{-a_j}\cos\left(\frac{2\pi\delta}{2^j}\right)\right).
\end{equation}
Generically, $\delta = 1/2$ results in the lowest probability of success, so
\begin{equation}
    p_s \geq \frac{1}{2}\prod_{j=1}^J\left(1+e^{-a_j}\cos\left(\frac{\pi}{2^j}\right)\right).
\end{equation}

Even in the noiseless case, the $1st$ term of this product will be $1/2$ resulting in a total probability less than $1/2$.
In order to amplify the success probability by taking multiple shots, the single shot probability must be greater than $1/2$.
This success probability can be boosted by discarding the $J$th bit, effectively rounding up or down based on the value of the $J$th bit.
Doing this decreases the precision of the outcome but increases the success probability by a factor of 2.
It is straightforward to include imperfect state overlap $\gamma<1$ as the dependence is strictly linear. 
\begin{equation}
    p_s(\gamma) \geq \frac{\gamma}{2}\prod_{j=2}^J\left(1+e^{-a_j}\cos\left(\frac{\pi}{2^j}\right)\right).
\end{equation}
In the noiseless case this product converges to $8\gamma/\pi^2$ as $J \rightarrow \infty$.

If $p_s(\gamma) > 0.5$, the success probability can be amplified arbitrarily high by taking multiple shots and accepting the most common result.
Taking $N_s$ shots, the total success probability using this strategy can be bounded as
\begin{equation}
    P_s > \prod_{k=\lceil N_s/2 \rceil}^{N_s} {N_s \choose k}\left( p_s(\gamma)\right)^{k}\left( 1-p_s(\gamma)\right)^{N_s-k}.
\end{equation}

\subsection{RPE subject to depolarizing noise}

RPE utilizes a series of Hadamard test circuits as shown in Fig.~\ref{fig:H_test}.
We first provide intuition for RPE in the absence of noise and assuming $\gamma = 1$, we will then relax those assumptions and show how to perform RPE with imperfect state overlap in the presence of depolarizing noise.
We define $z_j = \bar{x}_j+i\bar{y}_j$ where $x_j$ and $y_j$ are outputs when the $I/S^{\dag}$ gate is applied, respectively.
For input state $\ket{\psi_0}$, the expected value of $z_j$ is $E[z_j] = \bra{\psi}U^{2^{j-1}}\ket{\psi} = e^{-i\phi2^{j-1}}$.
The estimator for the $j$th generation is $\hat{\theta_j} = arg(z_j)$.
The estimate for $\phi$ is then defined as
\begin{equation}
    \hat{\phi} = \frac{\hat{\theta_j}}{2^{J-1}}.
\end{equation}
This estimate is only correct up to factors of $2\pi/2^{J-1}$ and thus requires the error on all previous generations to be below $\pi/2^{j-1}$. 
As shown in~\cite{li2023low}, we can use Hoeffding's inequality to ensure that this condition is met with probability $P_s$ if we take the appropriate number of samples $N_s$
\begin{equation}
    N_s = \biggl\lceil \frac{16}{3}\left( \log{\left( \frac{4}{1-P_s} \right)}+\log{(J-1)} \right) \biggr\rceil.
\end{equation}
In this case, with probability $P_s$ the error on $\hat{\phi}$ is
\begin{equation}
    \epsilon_{\hat{\phi}} = \frac{1}{2^{J-1}\sqrt{N_s}}.
\end{equation}

We will now include the effects of imperfect state overlap and depolarizing noise. 
Depolarizing noise inflates the variance however it does not bias the estimator of $\phi$. 
Imperfect state overlap causes the estimator to become biased. 
Together, their effect is to alter the error on $\hat{\phi}$ to
\begin{equation}
    \epsilon_{\hat{\phi}} = \frac{1}{2^{J-1}}\sqrt{\frac{e^{2a_J}}{N_s}+\eta^2(\gamma)}.
\end{equation}
The bias on any given generation $\eta_j(\gamma) = \theta_j - E[\hat{\theta_j}]$ can be upper bounded by brute force as
\begin{widetext}
\begin{equation}
    |\eta_j(\gamma)| < \text{atan2}\left[\sin{ \left(\sin^{-1}{ \left( \frac{(1-\gamma)^2}{\sqrt{2}\gamma} \right) } - \frac{\pi}{4}\right)}+\frac{1-\gamma}{\gamma},\cos{\left(\sin^{-1}{\left(\frac{(1-\gamma)^2}{\sqrt{2}\gamma}\right)} - \frac{\pi}{4}\right)}+\frac{1-\gamma}{\gamma}\right] - \sin^{-1}{\left(\frac{(1-\gamma)^2}{\sqrt{2}\gamma}\right)} + \frac{\pi}{4}.
    \label{eq:bias}
\end{equation}
\end{widetext}
The bias on the full protocol thus decreases with $j$, such that
\begin{equation}
    \eta = \frac{\eta_j}{2^{j-1}}.
\end{equation}
The protocol can still succeed provided that $\eta_j < \pi/2$ such that the bias does not cause the the error on any generation to exceed $\pi/2^{j-1}$.
This condition is met for $\gamma > 0.536$.

Depolarization does not add to the bias, however, it does increase the variance on each $\hat{\theta_j}$ by an amount $e^{2a_j}$.
While this could be mitigated by an equivalent increase in $N_s$, doing this results in $T_{tot}$ scaling exponentially with $\epsilon^{-1}$ and is therefore undesirable.
Instead, the QEC code distance is used to control the value of $e^{2a_j}$ such that only a modest increase in $N_s$ is necessary.
Accounting for the bias and increase in variance, Hoeffding's inequality can again be used to show to find the appropriate $N_s$ to succeed with probability $P_s$,
\begin{equation}
    N_s = \biggl\lceil \frac{4e^{2a_{J-1}}}{\frac{\sqrt{3}}{2}\gamma-\left( 1-\gamma \right)}\left( \log{\left( \frac{4}{1-P_s} \right)}+\log{(J-1)} \right) \biggr\rceil.
    \label{eq:rpeNs}
\end{equation}

As $\phi = \lambda t$, where $\lambda$ is the eigenvalue we wish to calculate, our estimate for $\lambda$ is simply
\begin{equation}
    \hat{\lambda} = \frac{\hat{\phi}}{t}.
\end{equation}
Therefore with probability $P_s$ the error on $\hat{\lambda}$ for $J$ generations of RPE is
\begin{equation}
    \epsilon_{J} \leq \frac{1}{t2^{J-1}}\sqrt{\frac{e^{2a_J}}{N_s}+\eta^2(\gamma)}.    
\end{equation}

\section{Circuit Constructions}
\label{app:circuit_constructions}

Here we show explicit circuit constructions for the controlled unitary evolution. 
The Trotterized unitary is expressed as sequence of Pauli rotations
\begin{equation}
    U = e^{-ig_1Z_1t}e^{-ig_2Z_2t}e^{-ig_3X_1X_2t}e^{-ig_4Y_1Y_2t}.
\end{equation}

Each rotation in the product can be implemented with a single Z rotation and Clifford gates.
The first two terms are already single qubit Z-rotations are implemented using the directional control shown in Fig.~\ref{fig:DC}.
The $X_1X_2$ rotation is performed by transforming the $Z_1$ operator into $X_1X_2$ via two H gates and a CNOT.
After the rotation is performed this transformation is undone.
Similar to the $X_1X_2$ rotation, the $Y_1Y_2$ rotation is performed by transforming $Z_1$ into $Y_1Y_2$.
All these pieces are put together in a first-order Trotter product.
We note that we calculated the difference between the eigenphase of the Trotterized unitary and exact unitary by hand to check that the Trotter error is not a limiting factor.
For the curious reader the Trotter error with $t = 0.5$ at.u. on the eigenphase itself was $0.0002$ rad, resulting in an error of $0.0004$ Ha.
All simulations were performed using a target error $\geq 10^{-3}$ and therefore were not impacted by Trotter error.
 
\onecolumngrid

\begin{figure}[ht]
    \centering
    \vspace{5em}
    \setlength{\tabcolsep}{0.9em}
    \begin{tabular}{cccc}
        \begin{subfigure}[c]{0.5\textwidth}
            \centering
            \Qcircuit @C=1em @R=1em {
            \lstick{\ket{+}} &  \ctrl{1} & \qw \\
            \lstick{\ket{\psi}} &  \gate{e^{-iPt}} & \qw }
        \end{subfigure}&
        \textbf{=} &
         &
        \begin{subfigure}[c]{0.45\textwidth}
            \centering
            \Qcircuit @C=1em @R=1em {
            \lstick{\ket{+}} &  \ctrl{1} & \qw &  \ctrl{1} & \qw\\
            \lstick{\ket{\psi}} &  \targ & \gate{e^{iPt/2}} &  \targ & \qw }
         \end{subfigure}
    \end{tabular}
    \caption{Implementation of a directionally controlled rotation operation where $P$ is a Pauli string.}
    \label{fig:DC}
\end{figure}
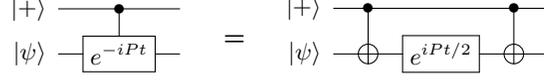

\begin{figure}[!ht]

    \centering
    \setlength{\tabcolsep}{0.9em}
    \begin{tabular}{cccc}
        \begin{subfigure}[c]{0.5\textwidth}
            \centering
            \Qcircuit @C=1em @R=.7em {
            \lstick{\ket{+}} & \ctrl{1} & \qw \\
            & \multigate{1}{e^{-ig_3X_1X_2t}}& \qw \\
            & \ghost{e^{-ig_3X_1X_2t}} & \qw }
        \end{subfigure}&
        \textbf{=} &
         &
        \begin{subfigure}[c]{0.45\textwidth}
            \centering
            \Qcircuit @C=1em @R=1em {
            \lstick{\ket{+}} & \qw & \qw & \ctrl{1} & \qw  & \ctrl{1} & \qw & \qw & \qw\\
            & \gate{H} & \targ & \targ & \gate{e^{ig_3Zt}}& \targ & \targ & \gate{H} & \qw\\
            & \gate{H} & \ctrl{-1} & \qw & \qw & \qw &  \ctrl{-1} & \gate{H} & \qw}
         \end{subfigure}
    \end{tabular}
    \caption{Directionally controlled-XX rotation}
\end{figure}
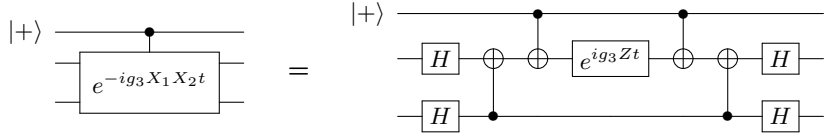

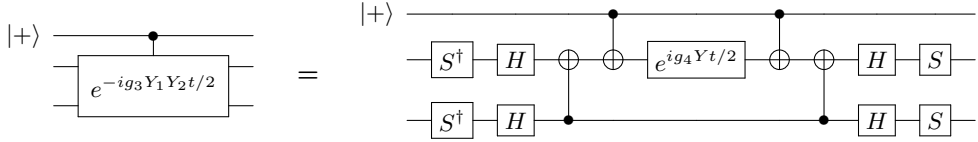
\begin{figure}[!ht]
    \centering

    \setlength{\tabcolsep}{0.9em}
    \begin{tabular}{cccc}
        \begin{subfigure}[c]{0.5\textwidth}
            \centering
            \Qcircuit @C=1em @R=.7em {
            \lstick{\ket{+}} & \ctrl{1} & \qw \\
            & \multigate{1}{e^{-ig_3Y_1Y_2t/2}}& \qw \\
            & \ghost{e^{-ig_3Y_1Y_2t/2}} & \qw }
        \end{subfigure}&
        \textbf{=} &
         &
        \begin{subfigure}[c]{0.45\textwidth}
            \centering
            \Qcircuit @C=1em @R=1em {
            \lstick{\ket{+}} & \qw & \qw & \qw & \ctrl{1} & \qw  & \ctrl{1} & \qw & \qw & \qw & \qw\\
            & \gate{S^{\dag}} & \gate{H} & \targ & \targ & \gate{e^{ig_4Yt/2}}& \targ & \targ & \gate{H} & \gate{S} & \qw\\
            & \gate{S^{\dag}} & \gate{H} & \ctrl{-1} & \qw & \qw & \qw &  \ctrl{-1} & \gate{H} & \gate{S} & \qw}
         \end{subfigure}
    \end{tabular}
    \caption{Directionally controlled-YY rotation}
\end{figure}    

\begin{figure}[!ht]
    \centering    
            \Qcircuit @C=1em @R=.7em {
            \lstick{\ket{+}} & \ctrl{1} & \qw  & \ctrl{1} & \ctrl{2} & \qw  & \ctrl{2} & \qw & \qw & \ctrl{1} & \qw  & \ctrl{1} & \qw & \qw & \qw & \qw & \qw & \ctrl{1} & \qw  & \ctrl{1} & \qw & \qw & \qw & \qw\\
            & \targ & \gate{e^{ig_1Zt/2}} & \targ & \qw  & \qw & \qw & \gate{H} & \targ & \targ & \gate{e^{ig_3Zt/2}}& \targ & \targ & \gate{H}  & \gate{S^{\dag}} & \gate{H} & \targ & \targ & \gate{e^{ig_4Yt/2}}& \targ & \targ & \gate{H} & \gate{S} & \qw\\
            & \qw  & \qw & \qw & \targ & \gate{e^{ig_2Zt/2}} & \targ & \gate{H} & \ctrl{-1} & \qw & \qw & \qw &  \ctrl{-1} & \gate{H} 
            & \gate{S^{\dag}} & \gate{H} & \ctrl{-1} & \qw & \qw & \qw &  \ctrl{-1} & \gate{H} & \gate{S} & \qw\\ }
    \caption{First-order Trotterized Hamiltonian simulation operator}
\vspace{2.5em}    
\end{figure}
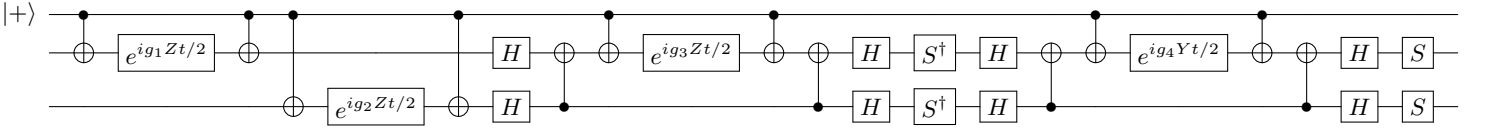

\twocolumngrid

\section{Numerical Tests}
\label{app:numerical_tests}
Here we show results from numerical simulations performed using qiskit.
For RPE, $N_s$ is selected according to~\ref{eq:rpeNs}.
For QCELS and MMQCELS,~\ref{eq:rpeNs} is used as a starting point, however at that $N_s$, these protocols did not reach the target error.
The $N_s$ was incremented until these protocols reached the desired precision.
For $\gamma=1$, the estimators for RPE, QCELS, and MMQCELS are unbiased, allowing trade-offs between $J$ and $N_s$.
For our simulations, $J$ was selected after $N_s$ had been chosen according to Hoeffding's inequality.
At $\gamma=1$ $J$ was then reduced by $\lceil \sqrt{N_s} \rceil$.

The protocols were compiled into Clifford+T circuits before simulating.
We assume a hardware error rate $p=10^{-3}$ and use~\ref{eq:aj} to select the minimum code distance $d$ such that the protocol would achieve the target error.
We introduced a depolarizing noise model using qiskit's AerSimulator.
The rate of the depolarizing noise was set according to~\ref{eq:aj}.
As we are performing resource estimates using Litinski-style architecture where all Clifford gates are commuted away we only apply our noise model only to the T gates.

The bias of an estimator can be defined as the error as $N_s \rightarrow \infty$.
In order to measure the bias on the estimators and compare to the upper bound of~\ref{eq:bias}, we ran simulations of RPE and QCELS at fixed $T_{max}$ while increasing $N_s$.
The bias is then found as the error on the estimate once the error remained constant rather than decreasing with increasing $N_s$. 
The bias for RPE and QCELS were found to be very similar for $\gamma=0.75$.
We note that as QCELS and MMQCELS use the same estimator, we expect their bias to be the same.
However, because of their reliance upon a classical nonlinear least-squares subroutine there might be subtle differences in practice due to differences in the landscapes.

\onecolumngrid

\vspace{1em}
\begin{figure}[!ht]
    \centering
    \includegraphics{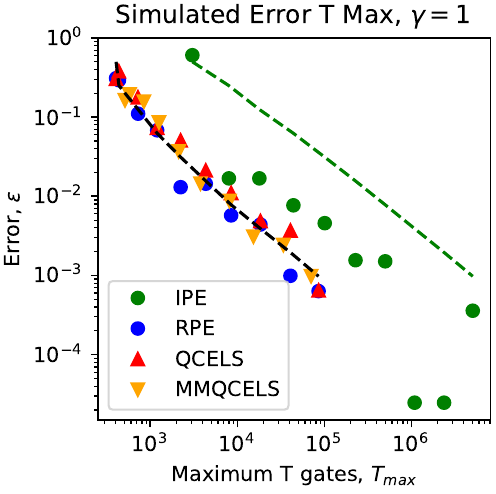}
    \includegraphics{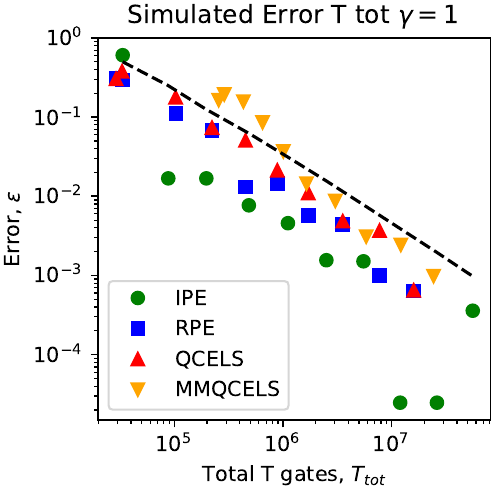}
    \caption{Error from numerical simulations at $\gamma=1$ for RPE, QCELS, and MMQCELS plotted against $T_{max}$ (left) and $T_{tot}$ (right). Dashed lines indicate the target error. As IPE has higher $T_{max}$ a separate green dashed line indicate the target error for IPE. Protocols were run 10 times each and the mean error of the estimates is shown.}
    \label{fig:simErrp1}
\end{figure}

\begin{figure}[!ht]
    \centering
    \includegraphics{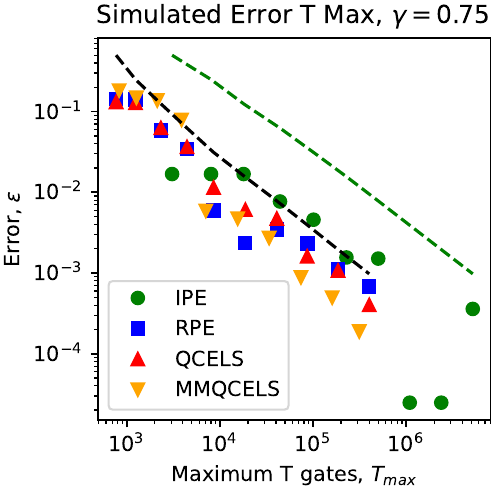}
    \includegraphics{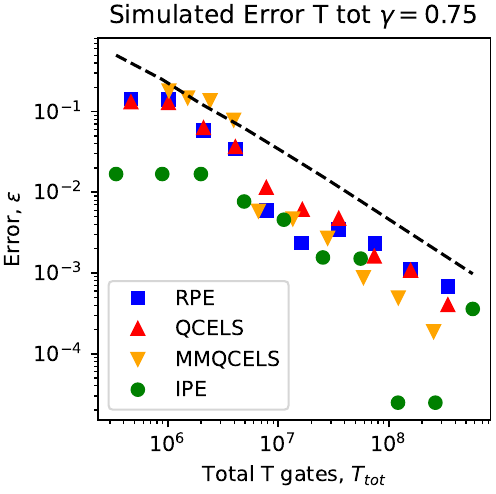}
    \caption{Error from numerical simulations at $\gamma=0.75$ for RPE, QCELS, and MMQCELS plotted against $T_{max}$ (left) and $T_{tot}$ (right). Dashed lines indicate the target error. As IPE has higher $T_{max}$ a separate green dashed line indicate the target error for IPE. Protocols were run 10 times each and the mean error of the estimates is shown.}
    \label{fig:simErrp1}
\end{figure}

\begin{figure}[!ht]
    \centering
    \includegraphics{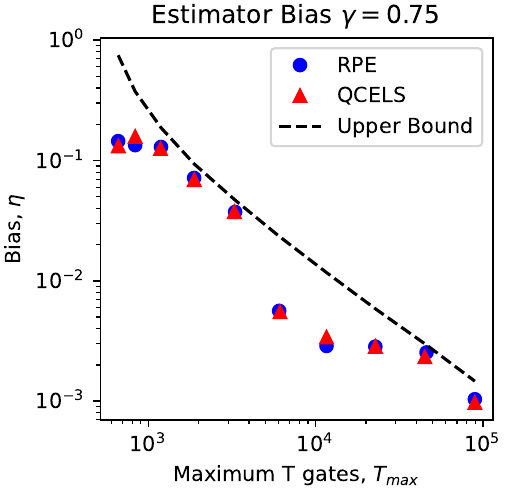}
    \caption{Estimator bias measured from numerical simulations. The dashed line shows the analytical upper bound.} 
    \label{fig:Simbias}
\end{figure}

\twocolumngrid

\clearpage
\bibliography{references.bib}

\begin{thebibliography}{69}%
\makeatletter
\providecommand \@ifxundefined [1]{%
 \@ifx{#1\undefined}
}%
\providecommand \@ifnum [1]{%
 \ifnum #1\expandafter \@firstoftwo
 \else \expandafter \@secondoftwo
 \fi
}%
\providecommand \@ifx [1]{%
 \ifx #1\expandafter \@firstoftwo
 \else \expandafter \@secondoftwo
 \fi
}%
\providecommand \natexlab [1]{#1}%
\providecommand \enquote  [1]{``#1''}%
\providecommand \bibnamefont  [1]{#1}%
\providecommand \bibfnamefont [1]{#1}%
\providecommand \citenamefont [1]{#1}%
\providecommand \href@noop [0]{\@secondoftwo}%
\providecommand \href [0]{\begingroup \@sanitize@url \@href}%
\providecommand \@href[1]{\@@startlink{#1}\@@href}%
\providecommand \@@href[1]{\endgroup#1\@@endlink}%
\providecommand \@sanitize@url [0]{\catcode `\\12\catcode `\$12\catcode `\&12\catcode `\#12\catcode `\^12\catcode `\_12\catcode `\%12\relax}%
\providecommand \@@startlink[1]{}%
\providecommand \@@endlink[0]{}%
\providecommand \url  [0]{\begingroup\@sanitize@url \@url }%
\providecommand \@url [1]{\endgroup\@href {#1}{\urlprefix }}%
\providecommand \urlprefix  [0]{URL }%
\providecommand \Eprint [0]{\href }%
\providecommand \doibase [0]{https://doi.org/}%
\providecommand \selectlanguage [0]{\@gobble}%
\providecommand \bibinfo  [0]{\@secondoftwo}%
\providecommand \bibfield  [0]{\@secondoftwo}%
\providecommand \translation [1]{[#1]}%
\providecommand \BibitemOpen [0]{}%
\providecommand \bibitemStop [0]{}%
\providecommand \bibitemNoStop [0]{.\EOS\space}%
\providecommand \EOS [0]{\spacefactor3000\relax}%
\providecommand \BibitemShut  [1]{\csname bibitem#1\endcsname}%
\let\auto@bib@innerbib\@empty
\bibitem [{\citenamefont {Ryan-Anderson}\ \emph {et~al.}(2021)\citenamefont {Ryan-Anderson}, \citenamefont {Bohnet}, \citenamefont {Lee}, \citenamefont {Gresh}, \citenamefont {Hankin}, \citenamefont {Gaebler}, \citenamefont {Francois}, \citenamefont {Chernoguzov}, \citenamefont {Lucchetti}, \citenamefont {Brown} \emph {et~al.}}]{ryan2021realization}%
  \BibitemOpen
  \bibfield  {author} {\bibinfo {author} {\bibfnamefont {C.}~\bibnamefont {Ryan-Anderson}}, \bibinfo {author} {\bibfnamefont {J.~G.}\ \bibnamefont {Bohnet}}, \bibinfo {author} {\bibfnamefont {K.}~\bibnamefont {Lee}}, \bibinfo {author} {\bibfnamefont {D.}~\bibnamefont {Gresh}}, \bibinfo {author} {\bibfnamefont {A.}~\bibnamefont {Hankin}}, \bibinfo {author} {\bibfnamefont {J.}~\bibnamefont {Gaebler}}, \bibinfo {author} {\bibfnamefont {D.}~\bibnamefont {Francois}}, \bibinfo {author} {\bibfnamefont {A.}~\bibnamefont {Chernoguzov}}, \bibinfo {author} {\bibfnamefont {D.}~\bibnamefont {Lucchetti}}, \bibinfo {author} {\bibfnamefont {N.~C.}\ \bibnamefont {Brown}}, \emph {et~al.},\ }\bibfield  {title} {\bibinfo {title} {Realization of real-time fault-tolerant quantum error correction},\ }\href {https://doi.org/10.1103/PhysRevX.11.041058} {\bibfield  {journal} {\bibinfo  {journal} {Physical Review X}\ }\textbf {\bibinfo {volume} {11}},\ \bibinfo {pages} {041058} (\bibinfo {year} {2021})}\BibitemShut {NoStop}%
\bibitem [{\citenamefont {{Google Quantum AI}}(2023)}]{google2023suppressing}%
  \BibitemOpen
  \bibfield  {author} {\bibinfo {author} {\bibnamefont {{Google Quantum AI}}},\ }\bibfield  {title} {\bibinfo {title} {Suppressing quantum errors by scaling a surface code logical qubit},\ }\href {https://doi.org/10.1038/s41586-022-05434-1} {\bibfield  {journal} {\bibinfo  {journal} {Nature}\ }\textbf {\bibinfo {volume} {614}},\ \bibinfo {pages} {676} (\bibinfo {year} {2023})}\BibitemShut {NoStop}%
\bibitem [{\citenamefont {Yamamoto}\ \emph {et~al.}(2023)\citenamefont {Yamamoto}, \citenamefont {Duffield}, \citenamefont {Kikuchi},\ and\ \citenamefont {Ramo}}]{yamamoto2023demonstrating}%
  \BibitemOpen
  \bibfield  {author} {\bibinfo {author} {\bibfnamefont {K.}~\bibnamefont {Yamamoto}}, \bibinfo {author} {\bibfnamefont {S.}~\bibnamefont {Duffield}}, \bibinfo {author} {\bibfnamefont {Y.}~\bibnamefont {Kikuchi}},\ and\ \bibinfo {author} {\bibfnamefont {D.~M.}\ \bibnamefont {Ramo}},\ }\href@noop {} {\bibinfo {title} {Demonstrating bayesian quantum phase estimation with quantum error detection}} (\bibinfo {year} {2023}),\ \Eprint {https://arxiv.org/abs/2306.16608} {arxiv:2306.16608 [quant-ph]} \BibitemShut {NoStop}%
\bibitem [{\citenamefont {Wang}\ \emph {et~al.}(2023)\citenamefont {Wang}, \citenamefont {Simsek}, \citenamefont {Gatterman}, \citenamefont {Gerber}, \citenamefont {Gilmore}, \citenamefont {Gresh}, \citenamefont {Hewitt}, \citenamefont {Horst}, \citenamefont {Matheny}, \citenamefont {Mengle}, \citenamefont {Neyenhuis},\ and\ \citenamefont {Criger}}]{wang2023faulttolerant}%
  \BibitemOpen
  \bibfield  {author} {\bibinfo {author} {\bibfnamefont {Y.}~\bibnamefont {Wang}}, \bibinfo {author} {\bibfnamefont {S.}~\bibnamefont {Simsek}}, \bibinfo {author} {\bibfnamefont {T.~M.}\ \bibnamefont {Gatterman}}, \bibinfo {author} {\bibfnamefont {J.~A.}\ \bibnamefont {Gerber}}, \bibinfo {author} {\bibfnamefont {K.}~\bibnamefont {Gilmore}}, \bibinfo {author} {\bibfnamefont {D.}~\bibnamefont {Gresh}}, \bibinfo {author} {\bibfnamefont {N.}~\bibnamefont {Hewitt}}, \bibinfo {author} {\bibfnamefont {C.~V.}\ \bibnamefont {Horst}}, \bibinfo {author} {\bibfnamefont {M.}~\bibnamefont {Matheny}}, \bibinfo {author} {\bibfnamefont {T.}~\bibnamefont {Mengle}}, \bibinfo {author} {\bibfnamefont {B.}~\bibnamefont {Neyenhuis}},\ and\ \bibinfo {author} {\bibfnamefont {B.}~\bibnamefont {Criger}},\ }\href@noop {} {\bibinfo {title} {Fault-tolerant one-bit addition with the smallest interesting colour code}} (\bibinfo {year} {2023}),\ \Eprint {https://arxiv.org/abs/2309.09893} {arXiv:2309.09893 [quant-ph]} \BibitemShut {NoStop}%
\bibitem [{\citenamefont {Bluvstein}\ \emph {et~al.}(2023)\citenamefont {Bluvstein}, \citenamefont {Evered}, \citenamefont {Geim}, \citenamefont {Li}, \citenamefont {Zhou}, \citenamefont {Manovitz}, \citenamefont {Ebadi}, \citenamefont {Cain}, \citenamefont {Kalinowski}, \citenamefont {Hangleiter} \emph {et~al.}}]{bluvstein2023logical}%
  \BibitemOpen
  \bibfield  {author} {\bibinfo {author} {\bibfnamefont {D.}~\bibnamefont {Bluvstein}}, \bibinfo {author} {\bibfnamefont {S.~J.}\ \bibnamefont {Evered}}, \bibinfo {author} {\bibfnamefont {A.~A.}\ \bibnamefont {Geim}}, \bibinfo {author} {\bibfnamefont {S.~H.}\ \bibnamefont {Li}}, \bibinfo {author} {\bibfnamefont {H.}~\bibnamefont {Zhou}}, \bibinfo {author} {\bibfnamefont {T.}~\bibnamefont {Manovitz}}, \bibinfo {author} {\bibfnamefont {S.}~\bibnamefont {Ebadi}}, \bibinfo {author} {\bibfnamefont {M.}~\bibnamefont {Cain}}, \bibinfo {author} {\bibfnamefont {M.}~\bibnamefont {Kalinowski}}, \bibinfo {author} {\bibfnamefont {D.}~\bibnamefont {Hangleiter}}, \emph {et~al.},\ }\bibfield  {title} {\bibinfo {title} {Logical quantum processor based on reconfigurable atom arrays},\ }\href {https://doi.org/10.1038/s41586-023-06927-3} {\bibfield  {journal} {\bibinfo  {journal} {Nature}\ ,\ \bibinfo {pages} {1}} (\bibinfo {year} {2023})}\BibitemShut {NoStop}%
\bibitem [{\citenamefont {Campbell}(2021)}]{campbell2021early}%
  \BibitemOpen
  \bibfield  {author} {\bibinfo {author} {\bibfnamefont {E.~T.}\ \bibnamefont {Campbell}},\ }\bibfield  {title} {\bibinfo {title} {Early fault-tolerant simulations of the hubbard model},\ }\href {https://doi.org/10.1088/2058-9565/ac3110} {\bibfield  {journal} {\bibinfo  {journal} {Quantum Science and Technology}\ }\textbf {\bibinfo {volume} {7}},\ \bibinfo {pages} {015007} (\bibinfo {year} {2021})}\BibitemShut {NoStop}%
\bibitem [{\citenamefont {Dong}\ \emph {et~al.}(2022)\citenamefont {Dong}, \citenamefont {Lin},\ and\ \citenamefont {Tong}}]{dong2022ground}%
  \BibitemOpen
  \bibfield  {author} {\bibinfo {author} {\bibfnamefont {Y.}~\bibnamefont {Dong}}, \bibinfo {author} {\bibfnamefont {L.}~\bibnamefont {Lin}},\ and\ \bibinfo {author} {\bibfnamefont {Y.}~\bibnamefont {Tong}},\ }\bibfield  {title} {\bibinfo {title} {Ground-state preparation and energy estimation on early fault-tolerant quantum computers via quantum eigenvalue transformation of unitary matrices},\ }\href {https://doi.org/10.1103/PRXQuantum.3.040305} {\bibfield  {journal} {\bibinfo  {journal} {PRX Quantum}\ }\textbf {\bibinfo {volume} {3}},\ \bibinfo {pages} {040305} (\bibinfo {year} {2022})}\BibitemShut {NoStop}%
\bibitem [{\citenamefont {Ni}\ \emph {et~al.}(2023)\citenamefont {Ni}, \citenamefont {Li},\ and\ \citenamefont {Ying}}]{ni2023low}%
  \BibitemOpen
  \bibfield  {author} {\bibinfo {author} {\bibfnamefont {H.}~\bibnamefont {Ni}}, \bibinfo {author} {\bibfnamefont {H.}~\bibnamefont {Li}},\ and\ \bibinfo {author} {\bibfnamefont {L.}~\bibnamefont {Ying}},\ }\bibfield  {title} {\bibinfo {title} {On low-depth algorithms for quantum phase estimation},\ }\bibfield  {journal} {\bibinfo  {journal} {arXiv preprint arXiv:2302.02454}\ }\href {https://doi.org/10.48550/arXiv.2302.02454} {10.48550/arXiv.2302.02454} (\bibinfo {year} {2023})\BibitemShut {NoStop}%
\bibitem [{\citenamefont {Li}\ \emph {et~al.}(2023)\citenamefont {Li}, \citenamefont {Ni},\ and\ \citenamefont {Ying}}]{li2023low}%
  \BibitemOpen
  \bibfield  {author} {\bibinfo {author} {\bibfnamefont {H.}~\bibnamefont {Li}}, \bibinfo {author} {\bibfnamefont {H.}~\bibnamefont {Ni}},\ and\ \bibinfo {author} {\bibfnamefont {L.}~\bibnamefont {Ying}},\ }\bibfield  {title} {\bibinfo {title} {On low-depth quantum algorithms for robust multiple-phase estimation},\ }\bibfield  {journal} {\bibinfo  {journal} {arXiv preprint arXiv:2303.08099}\ }\href {https://doi.org/10.48550/arXiv.2303.08099} {10.48550/arXiv.2303.08099} (\bibinfo {year} {2023})\BibitemShut {NoStop}%
\bibitem [{\citenamefont {Ding}\ and\ \citenamefont {Lin}(2023{\natexlab{a}})}]{ding2023even}%
  \BibitemOpen
  \bibfield  {author} {\bibinfo {author} {\bibfnamefont {Z.}~\bibnamefont {Ding}}\ and\ \bibinfo {author} {\bibfnamefont {L.}~\bibnamefont {Lin}},\ }\bibfield  {title} {\bibinfo {title} {Even shorter quantum circuit for phase estimation on early fault-tolerant quantum computers with applications to ground-state energy estimation},\ }\href {https://doi.org/10.1103/PRXQuantum.4.020331} {\bibfield  {journal} {\bibinfo  {journal} {PRX Quantum}\ }\textbf {\bibinfo {volume} {4}},\ \bibinfo {pages} {020331} (\bibinfo {year} {2023}{\natexlab{a}})}\BibitemShut {NoStop}%
\bibitem [{\citenamefont {Ding}\ and\ \citenamefont {Lin}(2023{\natexlab{b}})}]{ding2023simultaneous}%
  \BibitemOpen
  \bibfield  {author} {\bibinfo {author} {\bibfnamefont {Z.}~\bibnamefont {Ding}}\ and\ \bibinfo {author} {\bibfnamefont {L.}~\bibnamefont {Lin}},\ }\bibfield  {title} {\bibinfo {title} {Simultaneous estimation of multiple eigenvalues with short-depth quantum circuit on early fault-tolerant quantum computers},\ }\bibfield  {journal} {\bibinfo  {journal} {arXiv preprint arXiv:2303.05714}\ }\href {https://doi.org/10.48550/arXiv.2303.05714} {10.48550/arXiv.2303.05714} (\bibinfo {year} {2023}{\natexlab{b}})\BibitemShut {NoStop}%
\bibitem [{\citenamefont {Ding}\ \emph {et~al.}(2024)\citenamefont {Ding}, \citenamefont {Li}, \citenamefont {Lin}, \citenamefont {Ni}, \citenamefont {Ying},\ and\ \citenamefont {Zhang}}]{ding2024quantum}%
  \BibitemOpen
  \bibfield  {author} {\bibinfo {author} {\bibfnamefont {Z.}~\bibnamefont {Ding}}, \bibinfo {author} {\bibfnamefont {H.}~\bibnamefont {Li}}, \bibinfo {author} {\bibfnamefont {L.}~\bibnamefont {Lin}}, \bibinfo {author} {\bibfnamefont {H.}~\bibnamefont {Ni}}, \bibinfo {author} {\bibfnamefont {L.}~\bibnamefont {Ying}},\ and\ \bibinfo {author} {\bibfnamefont {R.}~\bibnamefont {Zhang}},\ }\href@noop {} {\bibinfo {title} {Quantum multiple eigenvalue gaussian filtered search: an efficient and versatile quantum phase estimation method}} (\bibinfo {year} {2024}),\ \Eprint {https://arxiv.org/abs/2402.01013} {arxiv:2402.01013 [quant-ph]} \BibitemShut {NoStop}%
\bibitem [{\citenamefont {Katabarwa}\ \emph {et~al.}(2023)\citenamefont {Katabarwa}, \citenamefont {Gratsea}, \citenamefont {Caesura},\ and\ \citenamefont {Johnson}}]{katabarwa2023early}%
  \BibitemOpen
  \bibfield  {author} {\bibinfo {author} {\bibfnamefont {A.}~\bibnamefont {Katabarwa}}, \bibinfo {author} {\bibfnamefont {K.}~\bibnamefont {Gratsea}}, \bibinfo {author} {\bibfnamefont {A.}~\bibnamefont {Caesura}},\ and\ \bibinfo {author} {\bibfnamefont {P.~D.}\ \bibnamefont {Johnson}},\ }\href@noop {} {\bibinfo {title} {Early fault-tolerant quantum computing}} (\bibinfo {year} {2023}),\ \Eprint {https://arxiv.org/abs/2311.14814} {arXiv:2311.14814 [quant-ph]} \BibitemShut {NoStop}%
\bibitem [{\citenamefont {Kitaev}(1995)}]{Kitaev1995QuantumMA}%
  \BibitemOpen
  \bibfield  {author} {\bibinfo {author} {\bibfnamefont {A.~Y.}\ \bibnamefont {Kitaev}},\ }\bibfield  {title} {\bibinfo {title} {Quantum measurements and the {Abelian} stabilizer problem},\ }\href {https://api.semanticscholar.org/CorpusID:17023060} {\bibfield  {journal} {\bibinfo  {journal} {Electron. Colloquium Comput. Complex.}\ }\textbf {\bibinfo {volume} {TR96}} (\bibinfo {year} {1995})}\BibitemShut {NoStop}%
\bibitem [{\citenamefont {Nielsen}\ and\ \citenamefont {Chuang}(2000)}]{mikeIke}%
  \BibitemOpen
  \bibfield  {author} {\bibinfo {author} {\bibfnamefont {M.~A.}\ \bibnamefont {Nielsen}}\ and\ \bibinfo {author} {\bibfnamefont {I.~L.}\ \bibnamefont {Chuang}},\ }\href@noop {} {\emph {\bibinfo {title} {Quantum Computation and Quantum Information}}}\ (\bibinfo  {publisher} {Cambridge University Press},\ \bibinfo {year} {2000})\BibitemShut {NoStop}%
\bibitem [{\citenamefont {Litinski}(2019{\natexlab{a}})}]{Litinski2019gameofsurfacecodes}%
  \BibitemOpen
  \bibfield  {author} {\bibinfo {author} {\bibfnamefont {D.}~\bibnamefont {Litinski}},\ }\bibfield  {title} {\bibinfo {title} {A game of surface codes: {L}arge-scale quantum computing with lattice surgery},\ }\href {https://doi.org/10.22331/q-2019-03-05-128} {\bibfield  {journal} {\bibinfo  {journal} {{Quantum}}\ }\textbf {\bibinfo {volume} {3}},\ \bibinfo {pages} {128} (\bibinfo {year} {2019}{\natexlab{a}})}\BibitemShut {NoStop}%
\bibitem [{\citenamefont {Shor}(1995)}]{shor1995scheme}%
  \BibitemOpen
  \bibfield  {author} {\bibinfo {author} {\bibfnamefont {P.~W.}\ \bibnamefont {Shor}},\ }\bibfield  {title} {\bibinfo {title} {Scheme for reducing decoherence in quantum computer memory},\ }\href {https://doi.org/10.1103/PhysRevA.52.R2493} {\bibfield  {journal} {\bibinfo  {journal} {Physical review A}\ }\textbf {\bibinfo {volume} {52}},\ \bibinfo {pages} {R2493} (\bibinfo {year} {1995})}\BibitemShut {NoStop}%
\bibitem [{\citenamefont {Calderbank}\ and\ \citenamefont {Shor}(1996)}]{calderbank1996good}%
  \BibitemOpen
  \bibfield  {author} {\bibinfo {author} {\bibfnamefont {A.~R.}\ \bibnamefont {Calderbank}}\ and\ \bibinfo {author} {\bibfnamefont {P.~W.}\ \bibnamefont {Shor}},\ }\bibfield  {title} {\bibinfo {title} {Good quantum error-correcting codes exist},\ }\href {https://doi.org/10.1103/PhysRevA.54.1098} {\bibfield  {journal} {\bibinfo  {journal} {Physical Review A}\ }\textbf {\bibinfo {volume} {54}},\ \bibinfo {pages} {1098} (\bibinfo {year} {1996})}\BibitemShut {NoStop}%
\bibitem [{\citenamefont {Shor}(1996)}]{shor1996fault}%
  \BibitemOpen
  \bibfield  {author} {\bibinfo {author} {\bibfnamefont {P.~W.}\ \bibnamefont {Shor}},\ }\bibfield  {title} {\bibinfo {title} {Fault-tolerant quantum computation},\ }in\ \href {https://doi.org/10.1109/SFCS.1996.548464} {\emph {\bibinfo {booktitle} {Proceedings of 37th conference on foundations of computer science}}}\ (\bibinfo {organization} {IEEE},\ \bibinfo {year} {1996})\ pp.\ \bibinfo {pages} {56--65}\BibitemShut {NoStop}%
\bibitem [{\citenamefont {Blunt}\ \emph {et~al.}(2023)\citenamefont {Blunt}, \citenamefont {Gehér},\ and\ \citenamefont {Moylett}}]{blunt2023compilation}%
  \BibitemOpen
  \bibfield  {author} {\bibinfo {author} {\bibfnamefont {N.~S.}\ \bibnamefont {Blunt}}, \bibinfo {author} {\bibfnamefont {G.~P.}\ \bibnamefont {Gehér}},\ and\ \bibinfo {author} {\bibfnamefont {A.~E.}\ \bibnamefont {Moylett}},\ }\href@noop {} {\bibinfo {title} {Compilation of a simple chemistry application to quantum error correction primitives}} (\bibinfo {year} {2023}),\ \Eprint {https://arxiv.org/abs/2307.03233} {arXiv:2307.03233 [quant-ph]} \BibitemShut {NoStop}%
\bibitem [{\citenamefont {Suzuki}\ \emph {et~al.}(2022)\citenamefont {Suzuki}, \citenamefont {Endo}, \citenamefont {Fujii},\ and\ \citenamefont {Tokunaga}}]{suzuki2022quantum}%
  \BibitemOpen
  \bibfield  {author} {\bibinfo {author} {\bibfnamefont {Y.}~\bibnamefont {Suzuki}}, \bibinfo {author} {\bibfnamefont {S.}~\bibnamefont {Endo}}, \bibinfo {author} {\bibfnamefont {K.}~\bibnamefont {Fujii}},\ and\ \bibinfo {author} {\bibfnamefont {Y.}~\bibnamefont {Tokunaga}},\ }\bibfield  {title} {\bibinfo {title} {Quantum error mitigation as a universal error reduction technique: Applications from the nisq to the fault-tolerant quantum computing eras},\ }\href {https://doi.org/10.1103/PRXQuantum.3.010345} {\bibfield  {journal} {\bibinfo  {journal} {PRX Quantum}\ }\textbf {\bibinfo {volume} {3}},\ \bibinfo {pages} {010345} (\bibinfo {year} {2022})}\BibitemShut {NoStop}%
\bibitem [{\citenamefont {Fowler}\ \emph {et~al.}(2012)\citenamefont {Fowler}, \citenamefont {Mariantoni}, \citenamefont {Martinis},\ and\ \citenamefont {Cleland}}]{fowler2012surface}%
  \BibitemOpen
  \bibfield  {author} {\bibinfo {author} {\bibfnamefont {A.~G.}\ \bibnamefont {Fowler}}, \bibinfo {author} {\bibfnamefont {M.}~\bibnamefont {Mariantoni}}, \bibinfo {author} {\bibfnamefont {J.~M.}\ \bibnamefont {Martinis}},\ and\ \bibinfo {author} {\bibfnamefont {A.~N.}\ \bibnamefont {Cleland}},\ }\bibfield  {title} {\bibinfo {title} {Surface codes: Towards practical large-scale quantum computation},\ }\href {https://doi.org/10.1103/PhysRevA.86.032324} {\bibfield  {journal} {\bibinfo  {journal} {Physical Review A}\ }\textbf {\bibinfo {volume} {86}},\ \bibinfo {pages} {032324} (\bibinfo {year} {2012})}\BibitemShut {NoStop}%
\bibitem [{\citenamefont {Kitaev}\ \emph {et~al.}(2002)\citenamefont {Kitaev}, \citenamefont {Shen},\ and\ \citenamefont {Vyalyi}}]{kitaev2002classical}%
  \BibitemOpen
  \bibfield  {author} {\bibinfo {author} {\bibfnamefont {A.~Y.}\ \bibnamefont {Kitaev}}, \bibinfo {author} {\bibfnamefont {A.}~\bibnamefont {Shen}},\ and\ \bibinfo {author} {\bibfnamefont {M.~N.}\ \bibnamefont {Vyalyi}},\ }\href@noop {} {\emph {\bibinfo {title} {Classical and quantum computation}}},\ \bibinfo {number} {47}\ (\bibinfo  {publisher} {American Mathematical Soc.},\ \bibinfo {year} {2002})\BibitemShut {NoStop}%
\bibitem [{\citenamefont {Clark}\ \emph {et~al.}(2009)\citenamefont {Clark}, \citenamefont {Metodi}, \citenamefont {Gasster},\ and\ \citenamefont {Brown}}]{clark2009resource}%
  \BibitemOpen
  \bibfield  {author} {\bibinfo {author} {\bibfnamefont {C.~R.}\ \bibnamefont {Clark}}, \bibinfo {author} {\bibfnamefont {T.~S.}\ \bibnamefont {Metodi}}, \bibinfo {author} {\bibfnamefont {S.~D.}\ \bibnamefont {Gasster}},\ and\ \bibinfo {author} {\bibfnamefont {K.~R.}\ \bibnamefont {Brown}},\ }\bibfield  {title} {\bibinfo {title} {Resource requirements for fault-tolerant quantum simulation: The ground state of the transverse {Ising} model},\ }\href {https://doi.org/10.1103/PhysRevA.79.062314} {\bibfield  {journal} {\bibinfo  {journal} {Physical Review A}\ }\textbf {\bibinfo {volume} {79}},\ \bibinfo {pages} {062314} (\bibinfo {year} {2009})}\BibitemShut {NoStop}%
\bibitem [{\citenamefont {Aspuru-Guzik}\ \emph {et~al.}(2005)\citenamefont {Aspuru-Guzik}, \citenamefont {Dutoi}, \citenamefont {Love},\ and\ \citenamefont {Head-Gordon}}]{aspuru2005simulated}%
  \BibitemOpen
  \bibfield  {author} {\bibinfo {author} {\bibfnamefont {A.}~\bibnamefont {Aspuru-Guzik}}, \bibinfo {author} {\bibfnamefont {A.~D.}\ \bibnamefont {Dutoi}}, \bibinfo {author} {\bibfnamefont {P.~J.}\ \bibnamefont {Love}},\ and\ \bibinfo {author} {\bibfnamefont {M.}~\bibnamefont {Head-Gordon}},\ }\bibfield  {title} {\bibinfo {title} {Simulated quantum computation of molecular energies},\ }\href {https://doi.org/10.1126/science.1113479} {\bibfield  {journal} {\bibinfo  {journal} {Science}\ }\textbf {\bibinfo {volume} {309}},\ \bibinfo {pages} {1704} (\bibinfo {year} {2005})}\BibitemShut {NoStop}%
\bibitem [{\citenamefont {Pathak}\ \emph {et~al.}(2023)\citenamefont {Pathak}, \citenamefont {Russo}, \citenamefont {Seritan},\ and\ \citenamefont {Baczewski}}]{pathak2023quantifying}%
  \BibitemOpen
  \bibfield  {author} {\bibinfo {author} {\bibfnamefont {S.}~\bibnamefont {Pathak}}, \bibinfo {author} {\bibfnamefont {A.}~\bibnamefont {Russo}}, \bibinfo {author} {\bibfnamefont {S.}~\bibnamefont {Seritan}},\ and\ \bibinfo {author} {\bibfnamefont {A.}~\bibnamefont {Baczewski}},\ }\bibfield  {title} {\bibinfo {title} {Quantifying {T}-gate-count improvements for ground-state-energy estimation with near-optimal state preparation},\ }\href {https://doi.org/10.1103/PhysRevA.107.L040601} {\bibfield  {journal} {\bibinfo  {journal} {Physical Review A}\ }\textbf {\bibinfo {volume} {107}},\ \bibinfo {pages} {L040601} (\bibinfo {year} {2023})}\BibitemShut {NoStop}%
\bibitem [{\citenamefont {Knill}\ \emph {et~al.}(2007)\citenamefont {Knill}, \citenamefont {Ortiz},\ and\ \citenamefont {Somma}}]{knill2007optimal}%
  \BibitemOpen
  \bibfield  {author} {\bibinfo {author} {\bibfnamefont {E.}~\bibnamefont {Knill}}, \bibinfo {author} {\bibfnamefont {G.}~\bibnamefont {Ortiz}},\ and\ \bibinfo {author} {\bibfnamefont {R.~D.}\ \bibnamefont {Somma}},\ }\bibfield  {title} {\bibinfo {title} {Optimal quantum measurements of expectation values of observables},\ }\href {https://doi.org/10.1103/PhysRevA.75.012328} {\bibfield  {journal} {\bibinfo  {journal} {Physical Review A}\ }\textbf {\bibinfo {volume} {75}},\ \bibinfo {pages} {012328} (\bibinfo {year} {2007})}\BibitemShut {NoStop}%
\bibitem [{\citenamefont {Harrow}\ \emph {et~al.}(2009)\citenamefont {Harrow}, \citenamefont {Hassidim},\ and\ \citenamefont {Lloyd}}]{harrow2009quantum}%
  \BibitemOpen
  \bibfield  {author} {\bibinfo {author} {\bibfnamefont {A.~W.}\ \bibnamefont {Harrow}}, \bibinfo {author} {\bibfnamefont {A.}~\bibnamefont {Hassidim}},\ and\ \bibinfo {author} {\bibfnamefont {S.}~\bibnamefont {Lloyd}},\ }\bibfield  {title} {\bibinfo {title} {Quantum algorithm for linear systems of equations},\ }\href {https://doi.org/10.1103/PhysRevLett.103.150502} {\bibfield  {journal} {\bibinfo  {journal} {Physical review letters}\ }\textbf {\bibinfo {volume} {103}},\ \bibinfo {pages} {150502} (\bibinfo {year} {2009})}\BibitemShut {NoStop}%
\bibitem [{\citenamefont {Temme}\ \emph {et~al.}(2011)\citenamefont {Temme}, \citenamefont {Osborne}, \citenamefont {Vollbrecht}, \citenamefont {Poulin},\ and\ \citenamefont {Verstraete}}]{temme2011quantum}%
  \BibitemOpen
  \bibfield  {author} {\bibinfo {author} {\bibfnamefont {K.}~\bibnamefont {Temme}}, \bibinfo {author} {\bibfnamefont {T.~J.}\ \bibnamefont {Osborne}}, \bibinfo {author} {\bibfnamefont {K.~G.}\ \bibnamefont {Vollbrecht}}, \bibinfo {author} {\bibfnamefont {D.}~\bibnamefont {Poulin}},\ and\ \bibinfo {author} {\bibfnamefont {F.}~\bibnamefont {Verstraete}},\ }\bibfield  {title} {\bibinfo {title} {Quantum metropolis sampling},\ }\href {https://doi.org/10.1038/nature09770} {\bibfield  {journal} {\bibinfo  {journal} {Nature}\ }\textbf {\bibinfo {volume} {471}},\ \bibinfo {pages} {87} (\bibinfo {year} {2011})}\BibitemShut {NoStop}%
\bibitem [{Note1()}]{Note1}%
  \BibitemOpen
  \bibinfo {note} {Arguably, the use of QPE in metrology~\cite {berry2001optimal,QMetrology,higgins2007entanglement,} and the calibration of gates on physical qubits~\cite {kimmel2015robust,rudinger2017experimental} is earlier, but does not require fault tolerance.}\BibitemShut {Stop}%
\bibitem [{\citenamefont {Berry}\ \emph {et~al.}(2001)\citenamefont {Berry}, \citenamefont {Wiseman},\ and\ \citenamefont {Breslin}}]{berry2001optimal}%
  \BibitemOpen
  \bibfield  {author} {\bibinfo {author} {\bibfnamefont {D.~W.}\ \bibnamefont {Berry}}, \bibinfo {author} {\bibfnamefont {H.}~\bibnamefont {Wiseman}},\ and\ \bibinfo {author} {\bibfnamefont {J.}~\bibnamefont {Breslin}},\ }\bibfield  {title} {\bibinfo {title} {Optimal input states and feedback for interferometric phase estimation},\ }\href {https://doi.org/10.1103/PhysRevA.63.053804} {\bibfield  {journal} {\bibinfo  {journal} {Physical Review A}\ }\textbf {\bibinfo {volume} {63}},\ \bibinfo {pages} {053804} (\bibinfo {year} {2001})}\BibitemShut {NoStop}%
\bibitem [{\citenamefont {Giovannetti}\ \emph {et~al.}(2006)\citenamefont {Giovannetti}, \citenamefont {Lloyd},\ and\ \citenamefont {Maccone}}]{QMetrology}%
  \BibitemOpen
  \bibfield  {author} {\bibinfo {author} {\bibfnamefont {V.}~\bibnamefont {Giovannetti}}, \bibinfo {author} {\bibfnamefont {S.}~\bibnamefont {Lloyd}},\ and\ \bibinfo {author} {\bibfnamefont {L.}~\bibnamefont {Maccone}},\ }\bibfield  {title} {\bibinfo {title} {Quantum metrology},\ }\href {https://doi.org/10.1103/PhysRevLett.96.010401} {\bibfield  {journal} {\bibinfo  {journal} {Phys. Rev. Lett.}\ }\textbf {\bibinfo {volume} {96}},\ \bibinfo {pages} {010401} (\bibinfo {year} {2006})}\BibitemShut {NoStop}%
\bibitem [{\citenamefont {Zhou}\ \emph {et~al.}(2018)\citenamefont {Zhou}, \citenamefont {Zhang}, \citenamefont {Preskill},\ and\ \citenamefont {Jiang}}]{Zhou_2018Hesienburg}%
  \BibitemOpen
  \bibfield  {author} {\bibinfo {author} {\bibfnamefont {S.}~\bibnamefont {Zhou}}, \bibinfo {author} {\bibfnamefont {M.}~\bibnamefont {Zhang}}, \bibinfo {author} {\bibfnamefont {J.}~\bibnamefont {Preskill}},\ and\ \bibinfo {author} {\bibfnamefont {L.}~\bibnamefont {Jiang}},\ }\bibfield  {title} {\bibinfo {title} {Achieving the {Heisenberg} limit in quantum metrology using quantum error correction},\ }\bibfield  {journal} {\bibinfo  {journal} {Nature Communications}\ }\textbf {\bibinfo {volume} {9}},\ \href {https://doi.org/10.1038/s41467-017-02510-3} {10.1038/s41467-017-02510-3} (\bibinfo {year} {2018})\BibitemShut {NoStop}%
\bibitem [{\citenamefont {Zwierz}\ \emph {et~al.}(2010)\citenamefont {Zwierz}, \citenamefont {P\'erez-Delgado},\ and\ \citenamefont {Kok}}]{OptimalityHeisenburg}%
  \BibitemOpen
  \bibfield  {author} {\bibinfo {author} {\bibfnamefont {M.}~\bibnamefont {Zwierz}}, \bibinfo {author} {\bibfnamefont {C.~A.}\ \bibnamefont {P\'erez-Delgado}},\ and\ \bibinfo {author} {\bibfnamefont {P.}~\bibnamefont {Kok}},\ }\bibfield  {title} {\bibinfo {title} {General optimality of the {Heisenberg} limit for quantum metrology},\ }\href {https://doi.org/10.1103/PhysRevLett.105.180402} {\bibfield  {journal} {\bibinfo  {journal} {Phys. Rev. Lett.}\ }\textbf {\bibinfo {volume} {105}},\ \bibinfo {pages} {180402} (\bibinfo {year} {2010})}\BibitemShut {NoStop}%
\bibitem [{\citenamefont {Braginskiĭ}\ and\ \citenamefont {Vorontsov}(1975)}]{SQL}%
  \BibitemOpen
  \bibfield  {author} {\bibinfo {author} {\bibfnamefont {V.~B.}\ \bibnamefont {Braginskiĭ}}\ and\ \bibinfo {author} {\bibfnamefont {Y.~I.}\ \bibnamefont {Vorontsov}},\ }\bibfield  {title} {\bibinfo {title} {Quantum-mechanical limitations in macroscopic experiments and modern experimental technique},\ }\bibfield  {journal} {\bibinfo  {journal} {Sov. Phys. Usp}\ }\textbf {\bibinfo {volume} {17}},\ \href {https://doi.org/10.1070/PU1975v017n05ABEH004362} {10.1070/PU1975v017n05ABEH004362} (\bibinfo {year} {1975})\BibitemShut {NoStop}%
\bibitem [{\citenamefont {Ross}\ and\ \citenamefont {Selinger}(2014)}]{ross2014optimal}%
  \BibitemOpen
  \bibfield  {author} {\bibinfo {author} {\bibfnamefont {N.~J.}\ \bibnamefont {Ross}}\ and\ \bibinfo {author} {\bibfnamefont {P.}~\bibnamefont {Selinger}},\ }\href@noop {} {\bibinfo {title} {Optimal ancilla-free {Clifford}+{T} approximation of z-rotations}} (\bibinfo {year} {2014}),\ \Eprint {https://arxiv.org/abs/1403.2975} {arXiv:1403.2975 [quant-ph]} \BibitemShut {NoStop}%
\bibitem [{\citenamefont {Bravyi}\ and\ \citenamefont {Kitaev}(2005)}]{kitaevUniversal2005}%
  \BibitemOpen
  \bibfield  {author} {\bibinfo {author} {\bibfnamefont {S.}~\bibnamefont {Bravyi}}\ and\ \bibinfo {author} {\bibfnamefont {A.}~\bibnamefont {Kitaev}},\ }\bibfield  {title} {\bibinfo {title} {Universal quantum computation with ideal {Clifford} gates and noisy ancillas},\ }\href {https://doi.org/10.1103/PhysRevA.71.022316} {\bibfield  {journal} {\bibinfo  {journal} {Phys. Rev. A}\ }\textbf {\bibinfo {volume} {71}},\ \bibinfo {pages} {022316} (\bibinfo {year} {2005})}\BibitemShut {NoStop}%
\bibitem [{\citenamefont {Litinski}(2019{\natexlab{b}})}]{Litinski2019magicstate}%
  \BibitemOpen
  \bibfield  {author} {\bibinfo {author} {\bibfnamefont {D.}~\bibnamefont {Litinski}},\ }\bibfield  {title} {\bibinfo {title} {Magic state distillation: {N}ot as costly as you think},\ }\href {https://doi.org/10.22331/q-2019-12-02-205} {\bibfield  {journal} {\bibinfo  {journal} {{Quantum}}\ }\textbf {\bibinfo {volume} {3}},\ \bibinfo {pages} {205} (\bibinfo {year} {2019}{\natexlab{b}})}\BibitemShut {NoStop}%
\bibitem [{\citenamefont {Lin}\ and\ \citenamefont {Tong}(2022)}]{lin2022heisenberg}%
  \BibitemOpen
  \bibfield  {author} {\bibinfo {author} {\bibfnamefont {L.}~\bibnamefont {Lin}}\ and\ \bibinfo {author} {\bibfnamefont {Y.}~\bibnamefont {Tong}},\ }\bibfield  {title} {\bibinfo {title} {Heisenberg-limited ground-state energy estimation for early fault-tolerant quantum computers},\ }\href {https://doi.org/10.1103/PRXQuantum.3.010318} {\bibfield  {journal} {\bibinfo  {journal} {PRX Quantum}\ }\textbf {\bibinfo {volume} {3}},\ \bibinfo {pages} {010318} (\bibinfo {year} {2022})}\BibitemShut {NoStop}%
\bibitem [{\citenamefont {Zhang}\ \emph {et~al.}(2022)\citenamefont {Zhang}, \citenamefont {Wang},\ and\ \citenamefont {Johnson}}]{zhang2022computing}%
  \BibitemOpen
  \bibfield  {author} {\bibinfo {author} {\bibfnamefont {R.}~\bibnamefont {Zhang}}, \bibinfo {author} {\bibfnamefont {G.}~\bibnamefont {Wang}},\ and\ \bibinfo {author} {\bibfnamefont {P.}~\bibnamefont {Johnson}},\ }\bibfield  {title} {\bibinfo {title} {Computing ground state properties with early fault-tolerant quantum computers},\ }\href {https://doi.org/10.22331/q-2022-07-11-761} {\bibfield  {journal} {\bibinfo  {journal} {Quantum}\ }\textbf {\bibinfo {volume} {6}},\ \bibinfo {pages} {761} (\bibinfo {year} {2022})}\BibitemShut {NoStop}%
\bibitem [{\citenamefont {Lu}\ \emph {et~al.}(2021)\citenamefont {Lu}, \citenamefont {Banuls},\ and\ \citenamefont {Cirac}}]{lu2021algorithms}%
  \BibitemOpen
  \bibfield  {author} {\bibinfo {author} {\bibfnamefont {S.}~\bibnamefont {Lu}}, \bibinfo {author} {\bibfnamefont {M.~C.}\ \bibnamefont {Banuls}},\ and\ \bibinfo {author} {\bibfnamefont {J.~I.}\ \bibnamefont {Cirac}},\ }\bibfield  {title} {\bibinfo {title} {Algorithms for quantum simulation at finite energies},\ }\href {https://doi.org/10.1103/PRXQuantum.2.020321} {\bibfield  {journal} {\bibinfo  {journal} {PRX Quantum}\ }\textbf {\bibinfo {volume} {2}},\ \bibinfo {pages} {020321} (\bibinfo {year} {2021})}\BibitemShut {NoStop}%
\bibitem [{\citenamefont {O’Brien}\ \emph {et~al.}(2021)\citenamefont {O’Brien}, \citenamefont {Polla}, \citenamefont {Rubin}, \citenamefont {Huggins}, \citenamefont {McArdle}, \citenamefont {Boixo}, \citenamefont {McClean},\ and\ \citenamefont {Babbush}}]{obrien2021error}%
  \BibitemOpen
  \bibfield  {author} {\bibinfo {author} {\bibfnamefont {T.~E.}\ \bibnamefont {O’Brien}}, \bibinfo {author} {\bibfnamefont {S.}~\bibnamefont {Polla}}, \bibinfo {author} {\bibfnamefont {N.~C.}\ \bibnamefont {Rubin}}, \bibinfo {author} {\bibfnamefont {W.~J.}\ \bibnamefont {Huggins}}, \bibinfo {author} {\bibfnamefont {S.}~\bibnamefont {McArdle}}, \bibinfo {author} {\bibfnamefont {S.}~\bibnamefont {Boixo}}, \bibinfo {author} {\bibfnamefont {J.~R.}\ \bibnamefont {McClean}},\ and\ \bibinfo {author} {\bibfnamefont {R.}~\bibnamefont {Babbush}},\ }\bibfield  {title} {\bibinfo {title} {Error mitigation via verified phase estimation},\ }\href {https://doi.org/10.1103/PRXQuantum.2.020317} {\bibfield  {journal} {\bibinfo  {journal} {PRX Quantum}\ }\textbf {\bibinfo {volume} {2}},\ \bibinfo {pages} {020317} (\bibinfo {year} {2021})}\BibitemShut {NoStop}%
\bibitem [{\citenamefont {Kimmel}\ \emph {et~al.}(2015)\citenamefont {Kimmel}, \citenamefont {Low},\ and\ \citenamefont {Yoder}}]{kimmel2015robust}%
  \BibitemOpen
  \bibfield  {author} {\bibinfo {author} {\bibfnamefont {S.}~\bibnamefont {Kimmel}}, \bibinfo {author} {\bibfnamefont {G.~H.}\ \bibnamefont {Low}},\ and\ \bibinfo {author} {\bibfnamefont {T.~J.}\ \bibnamefont {Yoder}},\ }\bibfield  {title} {\bibinfo {title} {Robust calibration of a universal single-qubit gate set via robust phase estimation},\ }\href {https://doi.org/10.1103/PhysRevA.92.062315} {\bibfield  {journal} {\bibinfo  {journal} {Physical Review A}\ }\textbf {\bibinfo {volume} {92}},\ \bibinfo {pages} {062315} (\bibinfo {year} {2015})}\BibitemShut {NoStop}%
\bibitem [{\citenamefont {Russo}\ \emph {et~al.}(2021{\natexlab{a}})\citenamefont {Russo}, \citenamefont {Rudinger}, \citenamefont {Morrison},\ and\ \citenamefont {Baczewski}}]{russo2021evaluating}%
  \BibitemOpen
  \bibfield  {author} {\bibinfo {author} {\bibfnamefont {A.~E.}\ \bibnamefont {Russo}}, \bibinfo {author} {\bibfnamefont {K.~M.}\ \bibnamefont {Rudinger}}, \bibinfo {author} {\bibfnamefont {B.~C.}\ \bibnamefont {Morrison}},\ and\ \bibinfo {author} {\bibfnamefont {A.~D.}\ \bibnamefont {Baczewski}},\ }\bibfield  {title} {\bibinfo {title} {Evaluating energy differences on a quantum computer with robust phase estimation},\ }\href {https://doi.org/10.1103/PhysRevLett.126.210501} {\bibfield  {journal} {\bibinfo  {journal} {Physical review letters}\ }\textbf {\bibinfo {volume} {126}},\ \bibinfo {pages} {210501} (\bibinfo {year} {2021}{\natexlab{a}})}\BibitemShut {NoStop}%
\bibitem [{\citenamefont {Ding}\ \emph {et~al.}(2023)\citenamefont {Ding}, \citenamefont {Dong}, \citenamefont {Tong},\ and\ \citenamefont {Lin}}]{ding2023robust}%
  \BibitemOpen
  \bibfield  {author} {\bibinfo {author} {\bibfnamefont {Z.}~\bibnamefont {Ding}}, \bibinfo {author} {\bibfnamefont {Y.}~\bibnamefont {Dong}}, \bibinfo {author} {\bibfnamefont {Y.}~\bibnamefont {Tong}},\ and\ \bibinfo {author} {\bibfnamefont {L.}~\bibnamefont {Lin}},\ }\bibfield  {title} {\bibinfo {title} {Robust ground-state energy estimation under depolarizing noise},\ }\bibfield  {journal} {\bibinfo  {journal} {arXiv preprint arXiv:2307.11257}\ }\href {https://doi.org/10.48550/arxiv.2307.11257} {10.48550/arxiv.2307.11257} (\bibinfo {year} {2023})\BibitemShut {NoStop}%
\bibitem [{\citenamefont {Dob\ifmmode \check{s}\else \v{s}\fi{}\'{\i}\ifmmode~\check{c}\else \v{c}\fi{}ek}\ \emph {et~al.}(2007)\citenamefont {Dob\ifmmode \check{s}\else \v{s}\fi{}\'{\i}\ifmmode~\check{c}\else \v{c}\fi{}ek}, \citenamefont {Johansson}, \citenamefont {Shumeiko},\ and\ \citenamefont {Wendin}}]{IPE}%
  \BibitemOpen
  \bibfield  {author} {\bibinfo {author} {\bibfnamefont {M.}~\bibnamefont {Dob\ifmmode \check{s}\else \v{s}\fi{}\'{\i}\ifmmode~\check{c}\else \v{c}\fi{}ek}}, \bibinfo {author} {\bibfnamefont {G.}~\bibnamefont {Johansson}}, \bibinfo {author} {\bibfnamefont {V.}~\bibnamefont {Shumeiko}},\ and\ \bibinfo {author} {\bibfnamefont {G.}~\bibnamefont {Wendin}},\ }\bibfield  {title} {\bibinfo {title} {Arbitrary accuracy iterative quantum phase estimation algorithm using a single ancillary qubit: A two-qubit benchmark},\ }\href {https://doi.org/10.1103/PhysRevA.76.030306} {\bibfield  {journal} {\bibinfo  {journal} {Phys. Rev. A}\ }\textbf {\bibinfo {volume} {76}},\ \bibinfo {pages} {030306} (\bibinfo {year} {2007})}\BibitemShut {NoStop}%
\bibitem [{\citenamefont {Griffiths}\ and\ \citenamefont {Niu}(1996)}]{PhysRevLett.76.3228}%
  \BibitemOpen
  \bibfield  {author} {\bibinfo {author} {\bibfnamefont {R.~B.}\ \bibnamefont {Griffiths}}\ and\ \bibinfo {author} {\bibfnamefont {C.-S.}\ \bibnamefont {Niu}},\ }\bibfield  {title} {\bibinfo {title} {Semiclassical {Fourier} transform for quantum computation},\ }\href {https://doi.org/10.1103/PhysRevLett.76.3228} {\bibfield  {journal} {\bibinfo  {journal} {Phys. Rev. Lett.}\ }\textbf {\bibinfo {volume} {76}},\ \bibinfo {pages} {3228} (\bibinfo {year} {1996})}\BibitemShut {NoStop}%
\bibitem [{\citenamefont {Sarovar}\ \emph {et~al.}(2020)\citenamefont {Sarovar}, \citenamefont {Proctor}, \citenamefont {Rudinger}, \citenamefont {Young}, \citenamefont {Nielsen},\ and\ \citenamefont {Blume-Kohout}}]{sarovar2020detecting}%
  \BibitemOpen
  \bibfield  {author} {\bibinfo {author} {\bibfnamefont {M.}~\bibnamefont {Sarovar}}, \bibinfo {author} {\bibfnamefont {T.}~\bibnamefont {Proctor}}, \bibinfo {author} {\bibfnamefont {K.}~\bibnamefont {Rudinger}}, \bibinfo {author} {\bibfnamefont {K.}~\bibnamefont {Young}}, \bibinfo {author} {\bibfnamefont {E.}~\bibnamefont {Nielsen}},\ and\ \bibinfo {author} {\bibfnamefont {R.}~\bibnamefont {Blume-Kohout}},\ }\bibfield  {title} {\bibinfo {title} {Detecting crosstalk errors in quantum information processors},\ }\href {https://doi.org/10.22331/q-2020-09-11-321} {\bibfield  {journal} {\bibinfo  {journal} {Quantum}\ }\textbf {\bibinfo {volume} {4}},\ \bibinfo {pages} {321} (\bibinfo {year} {2020})}\BibitemShut {NoStop}%
\bibitem [{Note2()}]{Note2}%
  \BibitemOpen
  \bibinfo {note} {In contrast to the optimal Heisenberg-limited scaling $N_s = \protect \mathcal {O}(\epsilon ^{-1})$.}\BibitemShut {Stop}%
\bibitem [{\citenamefont {Russo}\ \emph {et~al.}(2021{\natexlab{b}})\citenamefont {Russo}, \citenamefont {Kirby}, \citenamefont {Rudinger}, \citenamefont {Baczewski},\ and\ \citenamefont {Kimmel}}]{russo2021consistency}%
  \BibitemOpen
  \bibfield  {author} {\bibinfo {author} {\bibfnamefont {A.~E.}\ \bibnamefont {Russo}}, \bibinfo {author} {\bibfnamefont {W.~M.}\ \bibnamefont {Kirby}}, \bibinfo {author} {\bibfnamefont {K.~M.}\ \bibnamefont {Rudinger}}, \bibinfo {author} {\bibfnamefont {A.~D.}\ \bibnamefont {Baczewski}},\ and\ \bibinfo {author} {\bibfnamefont {S.}~\bibnamefont {Kimmel}},\ }\bibfield  {title} {\bibinfo {title} {Consistency testing for robust phase estimation},\ }\href {https://doi.org/10.1103/PhysRevA.103.042609} {\bibfield  {journal} {\bibinfo  {journal} {Physical Review A}\ }\textbf {\bibinfo {volume} {103}},\ \bibinfo {pages} {042609} (\bibinfo {year} {2021}{\natexlab{b}})}\BibitemShut {NoStop}%
\bibitem [{\citenamefont {Dennis}\ \emph {et~al.}(2002)\citenamefont {Dennis}, \citenamefont {Kitaev}, \citenamefont {Landahl},\ and\ \citenamefont {Preskill}}]{dennis2002topological}%
  \BibitemOpen
  \bibfield  {author} {\bibinfo {author} {\bibfnamefont {E.}~\bibnamefont {Dennis}}, \bibinfo {author} {\bibfnamefont {A.}~\bibnamefont {Kitaev}}, \bibinfo {author} {\bibfnamefont {A.}~\bibnamefont {Landahl}},\ and\ \bibinfo {author} {\bibfnamefont {J.}~\bibnamefont {Preskill}},\ }\bibfield  {title} {\bibinfo {title} {Topological quantum memory},\ }\href {https://doi.org/10.1063/1.1499754} {\bibfield  {journal} {\bibinfo  {journal} {Journal of Mathematical Physics}\ }\textbf {\bibinfo {volume} {43}},\ \bibinfo {pages} {4452} (\bibinfo {year} {2002})}\BibitemShut {NoStop}%
\bibitem [{\citenamefont {Eastin}\ and\ \citenamefont {Knill}(2009)}]{eastin2009restrictions}%
  \BibitemOpen
  \bibfield  {author} {\bibinfo {author} {\bibfnamefont {B.}~\bibnamefont {Eastin}}\ and\ \bibinfo {author} {\bibfnamefont {E.}~\bibnamefont {Knill}},\ }\bibfield  {title} {\bibinfo {title} {Restrictions on transversal encoded quantum gate sets},\ }\href {https://doi.org/10.1103/PhysRevLett.102.110502} {\bibfield  {journal} {\bibinfo  {journal} {Physical review letters}\ }\textbf {\bibinfo {volume} {102}},\ \bibinfo {pages} {110502} (\bibinfo {year} {2009})}\BibitemShut {NoStop}%
\bibitem [{\citenamefont {Dawson}\ and\ \citenamefont {Nielsen}(2005)}]{Dawson2005TheSA}%
  \BibitemOpen
  \bibfield  {author} {\bibinfo {author} {\bibfnamefont {C.~M.}\ \bibnamefont {Dawson}}\ and\ \bibinfo {author} {\bibfnamefont {M.~A.}\ \bibnamefont {Nielsen}},\ }\bibfield  {title} {\bibinfo {title} {The {Solovay-Kitaev} algorithm},\ }\href {https://api.semanticscholar.org/CorpusID:15371844} {\bibfield  {journal} {\bibinfo  {journal} {Quantum Inf. Comput.}\ }\textbf {\bibinfo {volume} {6}},\ \bibinfo {pages} {81} (\bibinfo {year} {2005})}\BibitemShut {NoStop}%
\bibitem [{\citenamefont {O'Malley}\ \emph {et~al.}(2016)\citenamefont {O'Malley}, \citenamefont {Babbush}, \citenamefont {Kivlichan}, \citenamefont {Romero}, \citenamefont {McClean}, \citenamefont {Barends}, \citenamefont {Kelly}, \citenamefont {Roushan}, \citenamefont {Tranter}, \citenamefont {Ding}, \citenamefont {Campbell}, \citenamefont {Chen}, \citenamefont {Chen}, \citenamefont {Chiaro}, \citenamefont {Dunsworth}, \citenamefont {Fowler}, \citenamefont {Jeffrey}, \citenamefont {Lucero}, \citenamefont {Megrant}, \citenamefont {Mutus}, \citenamefont {Neeley}, \citenamefont {Neill}, \citenamefont {Quintana}, \citenamefont {Sank}, \citenamefont {Vainsencher}, \citenamefont {Wenner}, \citenamefont {White}, \citenamefont {Coveney}, \citenamefont {Love}, \citenamefont {Neven}, \citenamefont {Aspuru-Guzik},\ and\ \citenamefont {Martinis}}]{O_Malley_2016molecules}%
  \BibitemOpen
  \bibfield  {author} {\bibinfo {author} {\bibfnamefont {P.}~\bibnamefont {O'Malley}}, \bibinfo {author} {\bibfnamefont {R.}~\bibnamefont {Babbush}}, \bibinfo {author} {\bibfnamefont {I.}~\bibnamefont {Kivlichan}}, \bibinfo {author} {\bibfnamefont {J.}~\bibnamefont {Romero}}, \bibinfo {author} {\bibfnamefont {J.}~\bibnamefont {McClean}}, \bibinfo {author} {\bibfnamefont {R.}~\bibnamefont {Barends}}, \bibinfo {author} {\bibfnamefont {J.}~\bibnamefont {Kelly}}, \bibinfo {author} {\bibfnamefont {P.}~\bibnamefont {Roushan}}, \bibinfo {author} {\bibfnamefont {A.}~\bibnamefont {Tranter}}, \bibinfo {author} {\bibfnamefont {N.}~\bibnamefont {Ding}}, \bibinfo {author} {\bibfnamefont {B.}~\bibnamefont {Campbell}}, \bibinfo {author} {\bibfnamefont {Y.}~\bibnamefont {Chen}}, \bibinfo {author} {\bibfnamefont {Z.}~\bibnamefont {Chen}}, \bibinfo {author} {\bibfnamefont {B.}~\bibnamefont {Chiaro}}, \bibinfo {author} {\bibfnamefont {A.}~\bibnamefont {Dunsworth}}, \bibinfo {author} {\bibfnamefont {A.}~\bibnamefont {Fowler}},
  \bibinfo {author} {\bibfnamefont {E.}~\bibnamefont {Jeffrey}}, \bibinfo {author} {\bibfnamefont {E.}~\bibnamefont {Lucero}}, \bibinfo {author} {\bibfnamefont {A.}~\bibnamefont {Megrant}}, \bibinfo {author} {\bibfnamefont {J.}~\bibnamefont {Mutus}}, \bibinfo {author} {\bibfnamefont {M.}~\bibnamefont {Neeley}}, \bibinfo {author} {\bibfnamefont {C.}~\bibnamefont {Neill}}, \bibinfo {author} {\bibfnamefont {C.}~\bibnamefont {Quintana}}, \bibinfo {author} {\bibfnamefont {D.}~\bibnamefont {Sank}}, \bibinfo {author} {\bibfnamefont {A.}~\bibnamefont {Vainsencher}}, \bibinfo {author} {\bibfnamefont {J.}~\bibnamefont {Wenner}}, \bibinfo {author} {\bibfnamefont {T.}~\bibnamefont {White}}, \bibinfo {author} {\bibfnamefont {P.}~\bibnamefont {Coveney}}, \bibinfo {author} {\bibfnamefont {P.}~\bibnamefont {Love}}, \bibinfo {author} {\bibfnamefont {H.}~\bibnamefont {Neven}}, \bibinfo {author} {\bibfnamefont {A.}~\bibnamefont {Aspuru-Guzik}},\ and\ \bibinfo {author} {\bibfnamefont {J.}~\bibnamefont {Martinis}},\ }\bibfield
  {title} {\bibinfo {title} {Scalable quantum simulation of molecular energies},\ }\bibfield  {journal} {\bibinfo  {journal} {Physical Review X}\ }\textbf {\bibinfo {volume} {6}},\ \href {https://doi.org/10.1103/physrevx.6.031007} {10.1103/physrevx.6.031007} (\bibinfo {year} {2016})\BibitemShut {NoStop}%
\bibitem [{\citenamefont {Wecker}\ \emph {et~al.}(2015)\citenamefont {Wecker}, \citenamefont {Hastings}, \citenamefont {Wiebe}, \citenamefont {Clark}, \citenamefont {Nayak},\ and\ \citenamefont {Troyer}}]{WeckerDirectControl}%
  \BibitemOpen
  \bibfield  {author} {\bibinfo {author} {\bibfnamefont {D.}~\bibnamefont {Wecker}}, \bibinfo {author} {\bibfnamefont {M.~B.}\ \bibnamefont {Hastings}}, \bibinfo {author} {\bibfnamefont {N.}~\bibnamefont {Wiebe}}, \bibinfo {author} {\bibfnamefont {B.~K.}\ \bibnamefont {Clark}}, \bibinfo {author} {\bibfnamefont {C.}~\bibnamefont {Nayak}},\ and\ \bibinfo {author} {\bibfnamefont {M.}~\bibnamefont {Troyer}},\ }\bibfield  {title} {\bibinfo {title} {Solving strongly correlated electron models on a quantum computer},\ }\href {https://doi.org/10.1103/PhysRevA.92.062318} {\bibfield  {journal} {\bibinfo  {journal} {Phys. Rev. A}\ }\textbf {\bibinfo {volume} {92}},\ \bibinfo {pages} {062318} (\bibinfo {year} {2015})}\BibitemShut {NoStop}%
\bibitem [{\citenamefont {Farhi}\ \emph {et~al.}(2000)\citenamefont {Farhi}, \citenamefont {Goldstone}, \citenamefont {Gutmann},\ and\ \citenamefont {Sipser}}]{farhi2000quantum}%
  \BibitemOpen
  \bibfield  {author} {\bibinfo {author} {\bibfnamefont {E.}~\bibnamefont {Farhi}}, \bibinfo {author} {\bibfnamefont {J.}~\bibnamefont {Goldstone}}, \bibinfo {author} {\bibfnamefont {S.}~\bibnamefont {Gutmann}},\ and\ \bibinfo {author} {\bibfnamefont {M.}~\bibnamefont {Sipser}},\ }\bibfield  {title} {\bibinfo {title} {Quantum computation by adiabatic evolution},\ }\bibfield  {journal} {\bibinfo  {journal} {arXiv preprint quant-ph/0001106}\ }\href {https://doi.org/10.48550/arXiv.quant-ph/0001106} {10.48550/arXiv.quant-ph/0001106} (\bibinfo {year} {2000})\BibitemShut {NoStop}%
\bibitem [{\citenamefont {Poulin}\ and\ \citenamefont {Wocjan}(2009)}]{poulin2009preparing}%
  \BibitemOpen
  \bibfield  {author} {\bibinfo {author} {\bibfnamefont {D.}~\bibnamefont {Poulin}}\ and\ \bibinfo {author} {\bibfnamefont {P.}~\bibnamefont {Wocjan}},\ }\bibfield  {title} {\bibinfo {title} {Preparing ground states of quantum many-body systems on a quantum computer},\ }\href {https://doi.org/10.1103/PhysRevLett.102.130503} {\bibfield  {journal} {\bibinfo  {journal} {Physical review letters}\ }\textbf {\bibinfo {volume} {102}},\ \bibinfo {pages} {130503} (\bibinfo {year} {2009})}\BibitemShut {NoStop}%
\bibitem [{\citenamefont {{Qiskit contributors}}(2023)}]{Qiskit}%
  \BibitemOpen
  \bibfield  {author} {\bibinfo {author} {\bibnamefont {{Qiskit contributors}}},\ }\href {https://doi.org/10.5281/zenodo.2573505} {\bibinfo {title} {Qiskit: An open-source framework for quantum computing}} (\bibinfo {year} {2023})\BibitemShut {NoStop}%
\bibitem [{\citenamefont {Fowler}\ and\ \citenamefont {Gidney}(2019)}]{fowler2019low}%
  \BibitemOpen
  \bibfield  {author} {\bibinfo {author} {\bibfnamefont {A.~G.}\ \bibnamefont {Fowler}}\ and\ \bibinfo {author} {\bibfnamefont {C.}~\bibnamefont {Gidney}},\ }\href@noop {} {\bibinfo {title} {Low overhead quantum computation using lattice surgery}} (\bibinfo {year} {2019}),\ \Eprint {https://arxiv.org/abs/1808.06709} {arXiv:1808.06709 [quant-ph]} \BibitemShut {NoStop}%
\bibitem [{\citenamefont {Fazio}\ \emph {et~al.}(2024)\citenamefont {Fazio}, \citenamefont {Harper},\ and\ \citenamefont {Bartlett}}]{fazio2024logical}%
  \BibitemOpen
  \bibfield  {author} {\bibinfo {author} {\bibfnamefont {N.}~\bibnamefont {Fazio}}, \bibinfo {author} {\bibfnamefont {R.}~\bibnamefont {Harper}},\ and\ \bibinfo {author} {\bibfnamefont {S.}~\bibnamefont {Bartlett}},\ }\href@noop {} {\bibinfo {title} {Logical noise bias in magic state injection}} (\bibinfo {year} {2024}),\ \Eprint {https://arxiv.org/abs/2401.10982} {arXiv:2401.10982 [quant-ph]} \BibitemShut {NoStop}%
\bibitem [{Note3()}]{Note3}%
  \BibitemOpen
  \bibinfo {note} {Source code for our simulations is available at \protect \href {https://github.com/adbacze/qpe4earlyftqc}{https://github.com/adbacze/qpe4earlyftqc}.}\BibitemShut {Stop}%
\bibitem [{\citenamefont {Rudinger}\ \emph {et~al.}(2019)\citenamefont {Rudinger}, \citenamefont {Proctor}, \citenamefont {Langharst}, \citenamefont {Sarovar}, \citenamefont {Young},\ and\ \citenamefont {Blume-Kohout}}]{rudinger2019probing}%
  \BibitemOpen
  \bibfield  {author} {\bibinfo {author} {\bibfnamefont {K.}~\bibnamefont {Rudinger}}, \bibinfo {author} {\bibfnamefont {T.}~\bibnamefont {Proctor}}, \bibinfo {author} {\bibfnamefont {D.}~\bibnamefont {Langharst}}, \bibinfo {author} {\bibfnamefont {M.}~\bibnamefont {Sarovar}}, \bibinfo {author} {\bibfnamefont {K.}~\bibnamefont {Young}},\ and\ \bibinfo {author} {\bibfnamefont {R.}~\bibnamefont {Blume-Kohout}},\ }\bibfield  {title} {\bibinfo {title} {Probing context-dependent errors in quantum processors},\ }\href {https://doi.org/10.1103/PhysRevX.9.021045} {\bibfield  {journal} {\bibinfo  {journal} {Physical Review X}\ }\textbf {\bibinfo {volume} {9}},\ \bibinfo {pages} {021045} (\bibinfo {year} {2019})}\BibitemShut {NoStop}%
\bibitem [{\citenamefont {Proctor}\ \emph {et~al.}(2020)\citenamefont {Proctor}, \citenamefont {Revelle}, \citenamefont {Nielsen}, \citenamefont {Rudinger}, \citenamefont {Lobser}, \citenamefont {Maunz}, \citenamefont {Blume-Kohout},\ and\ \citenamefont {Young}}]{proctor2020detecting}%
  \BibitemOpen
  \bibfield  {author} {\bibinfo {author} {\bibfnamefont {T.}~\bibnamefont {Proctor}}, \bibinfo {author} {\bibfnamefont {M.}~\bibnamefont {Revelle}}, \bibinfo {author} {\bibfnamefont {E.}~\bibnamefont {Nielsen}}, \bibinfo {author} {\bibfnamefont {K.}~\bibnamefont {Rudinger}}, \bibinfo {author} {\bibfnamefont {D.}~\bibnamefont {Lobser}}, \bibinfo {author} {\bibfnamefont {P.}~\bibnamefont {Maunz}}, \bibinfo {author} {\bibfnamefont {R.}~\bibnamefont {Blume-Kohout}},\ and\ \bibinfo {author} {\bibfnamefont {K.}~\bibnamefont {Young}},\ }\bibfield  {title} {\bibinfo {title} {Detecting and tracking drift in quantum information processors},\ }\href {https://doi.org/10.1038/s41467-020-19074-4} {\bibfield  {journal} {\bibinfo  {journal} {Nature communications}\ }\textbf {\bibinfo {volume} {11}},\ \bibinfo {pages} {5396} (\bibinfo {year} {2020})}\BibitemShut {NoStop}%
\bibitem [{\citenamefont {Wiebe}\ and\ \citenamefont {Granade}(2016)}]{wiebe2016efficient}%
  \BibitemOpen
  \bibfield  {author} {\bibinfo {author} {\bibfnamefont {N.}~\bibnamefont {Wiebe}}\ and\ \bibinfo {author} {\bibfnamefont {C.}~\bibnamefont {Granade}},\ }\bibfield  {title} {\bibinfo {title} {Efficient bayesian phase estimation},\ }\href {https://doi.org/10.1103/PhysRevLett.117.010503} {\bibfield  {journal} {\bibinfo  {journal} {Physical review letters}\ }\textbf {\bibinfo {volume} {117}},\ \bibinfo {pages} {010503} (\bibinfo {year} {2016})}\BibitemShut {NoStop}%
\bibitem [{\citenamefont {Low}\ and\ \citenamefont {Chuang}(2019)}]{Low2019Qubitization}%
  \BibitemOpen
  \bibfield  {author} {\bibinfo {author} {\bibfnamefont {G.~H.}\ \bibnamefont {Low}}\ and\ \bibinfo {author} {\bibfnamefont {I.~L.}\ \bibnamefont {Chuang}},\ }\bibfield  {title} {\bibinfo {title} {Hamiltonian simulation by qubitization},\ }\href {https://doi.org/10.22331/q-2019-07-12-163} {\bibfield  {journal} {\bibinfo  {journal} {{Quantum}}\ }\textbf {\bibinfo {volume} {3}},\ \bibinfo {pages} {163} (\bibinfo {year} {2019})}\BibitemShut {NoStop}%
\bibitem [{\citenamefont {Landahl}\ and\ \citenamefont {Ryan-Anderson}(2014)}]{landahl2014quantum}%
  \BibitemOpen
  \bibfield  {author} {\bibinfo {author} {\bibfnamefont {A.~J.}\ \bibnamefont {Landahl}}\ and\ \bibinfo {author} {\bibfnamefont {C.}~\bibnamefont {Ryan-Anderson}},\ }\href@noop {} {\bibinfo {title} {Quantum computing by color-code lattice surgery}} (\bibinfo {year} {2014}),\ \Eprint {https://arxiv.org/abs/1407.5103} {arXiv:1407.5103 [quant-ph]} \BibitemShut {NoStop}%
\bibitem [{\citenamefont {Liang}\ \emph {et~al.}(2023)\citenamefont {Liang}, \citenamefont {Zhou}, \citenamefont {Dalal},\ and\ \citenamefont {Johnson}}]{liang2023modeling}%
  \BibitemOpen
  \bibfield  {author} {\bibinfo {author} {\bibfnamefont {Q.}~\bibnamefont {Liang}}, \bibinfo {author} {\bibfnamefont {Y.}~\bibnamefont {Zhou}}, \bibinfo {author} {\bibfnamefont {A.}~\bibnamefont {Dalal}},\ and\ \bibinfo {author} {\bibfnamefont {P.~D.}\ \bibnamefont {Johnson}},\ }\href@noop {} {\bibinfo {title} {Modeling the performance of early fault-tolerant quantum algorithms}} (\bibinfo {year} {2023}),\ \Eprint {https://arxiv.org/abs/2306.17235} {arXiv:2306.17235 [quant-ph]} \BibitemShut {NoStop}%
\bibitem [{\citenamefont {Higgins}\ \emph {et~al.}(2007)\citenamefont {Higgins}, \citenamefont {Berry}, \citenamefont {Bartlett}, \citenamefont {Wiseman},\ and\ \citenamefont {Pryde}}]{higgins2007entanglement}%
  \BibitemOpen
  \bibfield  {author} {\bibinfo {author} {\bibfnamefont {B.~L.}\ \bibnamefont {Higgins}}, \bibinfo {author} {\bibfnamefont {D.~W.}\ \bibnamefont {Berry}}, \bibinfo {author} {\bibfnamefont {S.~D.}\ \bibnamefont {Bartlett}}, \bibinfo {author} {\bibfnamefont {H.~M.}\ \bibnamefont {Wiseman}},\ and\ \bibinfo {author} {\bibfnamefont {G.~J.}\ \bibnamefont {Pryde}},\ }\bibfield  {title} {\bibinfo {title} {Entanglement-free heisenberg-limited phase estimation},\ }\href {https://doi.org/10.1038/nature06257} {\bibfield  {journal} {\bibinfo  {journal} {Nature}\ }\textbf {\bibinfo {volume} {450}},\ \bibinfo {pages} {393} (\bibinfo {year} {2007})}\BibitemShut {NoStop}%
\bibitem [{\citenamefont {Rudinger}\ \emph {et~al.}(2017)\citenamefont {Rudinger}, \citenamefont {Kimmel}, \citenamefont {Lobser},\ and\ \citenamefont {Maunz}}]{rudinger2017experimental}%
  \BibitemOpen
  \bibfield  {author} {\bibinfo {author} {\bibfnamefont {K.}~\bibnamefont {Rudinger}}, \bibinfo {author} {\bibfnamefont {S.}~\bibnamefont {Kimmel}}, \bibinfo {author} {\bibfnamefont {D.}~\bibnamefont {Lobser}},\ and\ \bibinfo {author} {\bibfnamefont {P.}~\bibnamefont {Maunz}},\ }\bibfield  {title} {\bibinfo {title} {Experimental demonstration of a cheap and accurate phase estimation},\ }\href {https://doi.org/10.1103/PhysRevLett.118.190502} {\bibfield  {journal} {\bibinfo  {journal} {Physical review letters}\ }\textbf {\bibinfo {volume} {118}},\ \bibinfo {pages} {190502} (\bibinfo {year} {2017})}\BibitemShut {NoStop}%
\end{thebibliography}%

\end{document}